\documentclass[useAMS,usenatbib]{mn2e}
\usepackage{graphicx}
\usepackage{color}
\usepackage{ulem}
\def\apj{ApJ}
\def\apjs{ApJS}
\def\apjl{ApJL}
\def\aap{A\&A}
\def\aaps{A\&AS}
\def\aj{AJ}
\def\mnras{MNRAS}

\def\pasp{PASP}

\def\araa{Ann.Rev.Astron.Astrophys.}

\def\gal{Galaxies}
\def\assl{Astrophysics and Space Science Library}
\def\aspc{Astronomical Society of the Pacific}

\title[LBVs in DDO\,68~\#3 and PHL\,293B]
{Decade-long time-monitoring of candidate Luminous Blue Variable Stars in the two very metal-deficient compact dwarf galaxies DDO\,68 and PHL\,293B}  
\author[N. G. Guseva et al.]{N. G.\ Guseva$^{1}$,
T. X.\ Thuan$^{2,3}$ and Y. I.\ Izotov$^{1}$\\
                $^{1}$Bogolyubov Institute for Theoretical Physics,
                     Ukrainian National Academy of Sciences,
                     Metrologichna 14b, Kyiv 03143,  Ukraine,\\
                     nguseva@bitp.kiev.ua, yizotov@bitp.kiev.ua\\
                $^{2}$Astronomy Department, University of Virginia, 
                     P.O. Box 400325, Charlottesville, VA 22904-4325,\\
                     txt@virginia.edu\\
$^{3}$Institut d'Astrophysique de Paris (UMR 7095 CNRS \& Sorbonne Universit\'e) 98 bis Bd Arago, F-75014 Paris France\\
}
\begin{document}

\pagerange{\pageref{firstpage}--\pageref{lastpage}} \pubyear{2022}

\maketitle

\label{firstpage}

\begin{abstract}
 We have studied the spectral time variations of candidate 
luminous blue variable stars (cLBV) in two low-metallicity blue compact dwarf
galaxies,
DDO 68 and PHL 293B. 
The LBV in DDO 68, located in H~{\sc ii} region \#3, shows an outburst, with an increase of more than 1000 times in H$\alpha$ luminosity during the period  
2008--2010. 
The broad emission of the H~{\sc i} and He~{\sc i} lines display a P Cygni
profile, with a relatively constant terminal velocity of 
$\sim$800 km/s, 
reaching a maximum luminosity 
$L$(H$\alpha$) 
of $\sim$ 2$\times$10$^{38}$ erg/s,
with a FWHM of
$\sim$1000--1200 km/s.
  On the other hand, since the discovery of a cLBV in 2001 in PHL\,293B, the fluxes of the broad 
 components and the broad-to-narrow flux ratios of the H~{\sc i} and He~{\sc i} emission
 lines in this galaxy have remained nearly constant over 16 years, with small variations.
The luminosity of the broad H$\alpha$ component varies between $\sim$2$\times$10$^{38}$
erg/s and $\sim$10$^{39}$ erg/s, with the FWHM varying in the range $\sim$500--1500 km/s.
   Unusually persistent P Cygni features 
   are clearly visible until the end of 2020 despite 
a decrease of the broad-to-narrow flux ratio in the most recent years. 
A terminal velocity of $\sim$800 km/s is measured from the P Cygni profile, similar to 
the one in DDO\,68, although the latter is 3.7 more metal-deficient than PHL\,293B.
The relative constancy of the broad H$\alpha$ luminosity in 
PHL\,293B suggests that it is  
due to a long-lived stellar transient of type LBV/SN IIn.  
\end{abstract}


\section{Introduction}\label{sec:INT}

 During their evolution, massive stars are known to go through a very short and high luminosity important transitional phase, called the luminous blue variable (LBV) phase. During this phase, LBVs undergo significant variations in photometric magnitudes and spectral features, characterized in particular by the appearance of broad components in the hydrogen and helium emission lines and in some heavy
element ion lines in the  UV and optical ranges.
    The broad emission in the LBVs may be due to sharp eruptions of the 
massive stars, reaching a total mass loss of up to $\sim$30 -- 100 M$_\odot$.
    It can also be caused by stellar winds and
expanding dense circumstellar envelopes. In these cases, H$\alpha$ luminosities
are in the range of 10$^{36}$ -- 10$^{39}$ ergs s$^{-1}$ \citep{IzTG07}.
     The mass loss rate of hydrogen-rich layers 
through stellar winds is $\sim$10$^{-6}$ -- $\sim$10$^{-3}$M$_\odot$ yr$^{-1}$
\citep{HumphreysDavidson1994,Smith1994,Drissen1997,Drissen2001}.
    However, very strong broad emission (FWHM $>$ 1000 km s$^{-1}$) can also be  present in the spectra of objects other than massive stars, such as Type IIn Supernovae (SNe) and Active Galactic Nuclei (AGNs). In these objects, 
the luminosities of the broad H$\alpha$ component are larger and can reach values up to
10$^{40}$ -- 10$^{42}$ ergs s$^{-1}$  
\citep{IzTG07,Sobral2020,Kokubo2021,Burke2021}.
 
LBVs frequently show recurring eruptive events through various evolution phases, during their transition from young massive
main-sequence stars to WR stars, SN explosions or massive black holes (BHs).
It is believed that stars with masses greater than 20 -- 30 M$_\odot$ and
luminosities 
$L$ $\sim$ 10$^{3}$--10$^{6}$$L_\odot$
go through the LBV phase  \citep{Crowther2007,Solovyeva2020}.
   Among all types of variable stars, only LBV stars show significant
variability both in photometric brightness and spectroscopic features:
rapidly amplified broad emission and blueward absorption lines, strongly enhanced
continuum that becomes bluer in the UV and optical spectra.

To date, about a few hundreds LBVs and candidate LBVs (cLBVs) are known
to show irregular cyclic quasi-periodic brightness variations of $\sim$0.5 - 2 mag 
on timescales from several years to decades. They are called 
S Dor LBVs \citep[see e.g. ][]{Massey2000,Humphreys2013,Humphreys2017,Humphreys2019,Grassitelli2020,Weis2020}.
On the other hand, there exists a 
tiny number of LBV stars which show giant eruptions, with
amplitudes more than 2.5 - 3 mag, on timescales of up to thousands years   
\citep{DavidsonHumphreys1997,Smith2011,Vink2012,Weis2020}.  Well-known
prototypes of this category are  $\eta$ Carinae and P Cygni with luminosities
$\sim$10$^{40}$ ergs s$^{-1}$ \citep{Lamers1983,Davidson1999}.
   In some cases, the peak luminosity during the outbursts can reach
$\sim$10$^{42}$ ergs s$^{-1}$ \citep{Kokubo2021}.
Nearly all these known luminous LBVs are either in our Galaxy or in nearby galaxies.

  High intensity broad and very broad components of emission lines, with
P Cygni profiles, have also been 
observed in the integral spectra of 
star-forming galaxies (SFGs) underlying their 
strong narrow emission lines produced in H~{\sc ii} regions
\citep[see e.g. ][]{SCP99,GIT00,Lei01,IzT08,Guseva2012}. 
  The most prominent spectral features in SFGs with LBVs are broad
  components of hydrogen and often helium lines with blueward absorption,
  and Fe {\sc ii} emission. 

To understand the physical mechanism responsible for the broad emission, 
time-monitoring of the broad spectral features is necessary. The reason is that, as said previously, broad emission occurs not only 
in LBV spectra but also in those of SNe and AGNs.  
It is difficult to distinguish between these different 
possibilities without a long-term time monitoring of the broad features. 
    It has now been established that a significant number of the objects detected in
supernova surveys are not true supernovae, but belong to a category of objects
called ``supernova impostors''. 
    Ordinary LBVs of the S Dor type or LBVs with giant eruptions, like $\eta$ Car at maximum
luminosity, appear among these "impostors". 
Despite the fact that many stars with LBV features have been
discovered \citep{Weis2020}, only a few dozen of them are confirmed as genuine Galactic
and extragalactic LBVs \citep{Wofford2020}.
The remaining are cLBV which require time-monitoring to confirm their true nature. 
A genuine LBV would show a
significant enhancement of the spectral and
photometric features on a time scale of tens of years, followed by the disappearance of
these features. This disappearance is necessary to rule out SNe, AGNs or other physical mechanisms \citep{Kokubo2021}. 

In addition, it is of particular interest to understand how LBV evolution in SFGs depends on the
properties of the host galaxy,  
such as gas metallicity, interstellar medium density, star formation rate (SFR)
and specific SFR (sSFR). Only very few LBVs are known up to date in
metal-poor SFGs with strong star-forming activity
 \citep[see e.g. ][and references therein]{Weis2020}.
We discuss in this paper 
the time monitoring over two decades of 
the photometric and spectroscopic properties of cLBV stars in two extremely
metal-deficient dwarf star-forming galaxies (SFG), DDO 68 (located in H~{\sc ii} region \#3) with 12 + log (O/H) = 7.15 and PHL 293B with 12 + log (O/H) = 7.72. These two SFGs are the lowest-metallicity galaxies where LBV stars have been detected, allowing the study of the LBV phenomenon in the extremely low metallicity regime, and shedding light of the evolution of the first generation of massive stars possibly born from primordial gas.

The paper is structured as follows: the LBT/MODS and 3.5m APO observations
and data reduction are described in Sect.~\ref{sec:OBS}.
In Sect.~\ref{sec:Results} we present the study of multi-epoch optical spectra
of DDO\,68~\#3 and PHL\,293B.
Finally, in Sect.~\ref{sec:conclusion} we summarize our main results.

\begin{table}
\caption{Journal of observations  \label{tab6}}
\begin{tabular}{lcccc} \hline
Date   && exposure & slit width & airmass \\
       && (seconds)& (arcsec)   &          \\
\hline
\multicolumn{5}{c}{DDO\,68~\#3,  3.5m APO} \\
2008-10-27 && 2700 & 1.5 & 1.24 \\
2008-11-06 && 2400 & 1.5 & 1.17 \\
2009-02-22 && 1800 & 2.0 & 1.53 \\
2009-11-19 && 1800 & 2.0 & 1.18 \\
2010-02-06 && 2700 & 1.5 & 1.36 \\
2010-03-20 && 1800 & 1.5 & 1.16 \\
2010-10.31 && 1500 & 1.5 & 1.13 \\
2012-02-15 && 2700 & 1.5 & 1.20 \\
2012-05-16 && 1800 & 1.5 & 1.27 \\
2013-06-01 &&  986 & 1.5 & 1.42 \\
2016-04-09 && 1800 & 0.9 & 1.51 \\
2018-04-07 && 1860 & 1.5 & 1.06 \\
\multicolumn{5}{c}{PHL\,293B, 3.5m APO} \\
2010-10-06 && 1800 & 1.5 & 1.56 \\
2014-11.17 && 2700 & 0.9 & 1.19 \\
2015-11-06 && 1800 & 0.9 & 1.21 \\
2017-12-15 && 1800 & 0.9 & 1.27 \\
\multicolumn{5}{c}{PHL\,293B, 2$\times$8.4m LBT/MODS} \\
2020-11-18 && 2700 & 1.2 & 1.19 \\  
\hline
  \end{tabular}



  \end{table}

\begin{table*}
\caption{Parameters of H$\alpha$ narrow and broad components in DDO\,68~\#3 spectra, observed with the 3.5m APO telescope and presented in Fig.~\ref{fig6}  \label{tab7-DDO-G}}
\begin{tabular}{lccccccc} \hline

Date & $F_{nar}$$^{\rm a}$ &FWHM$_{nar}$$^{\rm b}$ &FWHM$^{\rm c}$ & $F_{br}$$^{\rm a}$ &FWHM$_{br}$$^{\rm b}$ & $v$$_{\rm term}$$^{\rm d}$  &$L_{br}$$^{\rm e}$ \\  
&&&([O~{\sc iii}]4959)&&&& \\
\hline
2008-10-27 &    64.26  & 115(2.51) & 2.42 & 84.01  & 1072   & 706    & 16.2 \\
2008-11-06 &    59.32  & 79(1.74)  & 2.00 & 66.86  & 1029   & 722    & 12.9 \\
2009-02-22 &    47.13  & 88(1.92)  & 2.33 & 74.41  & 1006   & 764    & 14.4 \\
2009-11-19 &    56.61  & 87(1.91)  & 2.16 & 93.66  & 1110   & 808    & 18.1 \\
2010-02-06 &    27.58  & 78(1.72)  & 1.94 & 30.94  & 1251   & 749    & 6.0 \\ 
2010-03-20 &    34.85  & 78(1.72)  & 2.15 & 49.48  & 1107   & 819    & 9.6 \\
2010-10-31 &    56.30  & 78(1.71)  & 2.55 & 73.84  & 1130   & 871    & 14.3 \\
2012-02-15 &    48.22  & 69(1.52)  & 1.86 & 29.10  & 742    &...     & 5.6 \\
2012-05-16 &    73.16  & 78(1.72) & 1.80 & 31.87  & 734     &...     & 6.2 \\
2013-06-01 &   45.34   & 74(1.61) & 1.61  & 35.82   & 560   &...     & 6.9 \\
2016-04-09 &   13.43   & 63(1.39) & 1.54  & 2.44    & 326   &...     & 0.5 \\
2018-04-07 &   8.48    & 86(1.87) & 1.63  &...      &  ...  &  ...   & 0.0 \\ \\
\hline
  \end{tabular}
\begin{tabular}{l}
$^{\rm a}$Fluxes of H$\alpha$ narrow ($F_{nar}$) and broad ($F_{br}$) components in units of 10$^{-16}$ erg s$^{-1}$ cm$^{-2}$. \\
  $^{\rm b}$Full widths at half maximum (FWHM) of H$\alpha$ narrow and broad  components in units of km s$^{-1}$, and in \\
  angstroms (in brackets).\\
$^{\rm c}$FWHM of [O {\sc iii}]$\lambda$4959\AA\ emission line in units of angstroms. \\
  $^{\rm d}$Terminal velocity in units of km s$^{-1}$. \\ 
  $^{\rm e}$Luminosity of H$\alpha$ broad component in units of 10$^{37}$ erg s$^{-1}$, calculated with a distance $D$=12.71 Mpc 
  \\ which is a mean 
  from determinations by  \citet{Cannon2014,Sacchi2016,Makarov2017}. \\
  
  \end{tabular}
\end{table*}

\begin{table*}
\caption{Parameters of H$\alpha$ narrow and broad components in PHL\,293B spectra, presented in Fig.~\ref{fig7}  \label{tab7-PHL-G}}
\begin{tabular}{llrccccccccc} \hline

  Date & tel.$^{\rm a}$ & $F^{\rm b}$ &FWHM$^{\rm c}$ &FWHM$^{\rm d}$ &$F^{\rm b}$ &FWHM$^{\rm c}$ &
$F^{\rm b}$ &FWHM$^{\rm c}$ & $F^{\rm e}$   & $v$$_{\rm term}$$^{\rm f}$ & $L^{\rm g}$ \\  
&&{\sl nar}&{\sl nar}&4959&{\sl br}&{\sl br}&{\sl v.br}&{\sl v.br}&{\sl sum br} &{\sl br} & {\sl sum br} \\
\hline
2001-08-22 & SDSS   & 238.02 &235(5.1)& 3.8 & 63.74 & 1379& 7.86 & 2602 & 71.60  & 938 &  44.2 \\
2010-10-06 & APO    & 261.50 & 71(1.6)& 2.2 & 27.13 & 181  &42.99 & 999 & 70.12  & 702 &  43.2 \\
2014-11-17 & APO    & 71.04  & 66(1.4)& 1.6 & 11.77 & 162  &12.62 & 627 & 24.39  &...  &  15.0 \\
2015-11-06 & APO    & 471.20 & 74(1.6)& 1.6 & 45.96 & 180  &95.08 & 705 &141.04  &...  &  87.0 \\
2017-12-15 & APO    & 91.16  & 69(1.5)& 1.6 & 9.64  & 165  &15.02 & 903 & 24.66  &...  &  15.2 \\
2020-11-18 & LBT    & 401.30 &280(6.1)& 4.0 & 23.08 & 504 &11.64 & 1320 & 34.72  &...  &  21.4 \\ \\
\hline
  \end{tabular}
\begin{tabular}{l}
$^{\rm a}$telescopes: 2.5m APO (labelled as SDSS), 3.5m APO (labelled as APO), 2$\times$8.4m LBT (labelled as LBT).  \\
  $^{\rm b}$Flux of narrow ($F_{nar}$), broad ($F_{br}$) and very broad ($F_{v.br}$) components of H$\alpha$ in units of 10$^{-16}$ erg s$^{-1}$ cm$^{-2}$. \\
  $^{\rm c}$Full width at half maximum (FWHM) of narrow, broad and very broad components of H$\alpha$ in km s$^{-1}$ and in angstroms\\
  (in brackets).\\
$^{\rm d}$FWHM of the [O~{\sc iii}]$\lambda$4959\AA\ emission line in angstroms. \\
$^{\rm e}$Flux of the whole broad bump (i.e. sum of the broad and very broad components) in units of 10$^{-16}$ erg s$^{-1}$ cm$^{-2}$. \\
  $^{\rm f}$Terminal velocity in units of km s$^{-1}$. \\ 
$^{\rm g}$Luminosity of the whole broad bump (i.e. the sum of broad and very broad components) in 10$^{37}$ erg s$^{-1}$, calculated with the \\
distance $D$=22.7 Mpc of PHL\,293B, the same as that adopted by \citet{IzT09}.  \\

  \end{tabular}
  \end{table*}

\begin{figure*}
\hbox{
\includegraphics[angle=-90,width=0.98\linewidth]{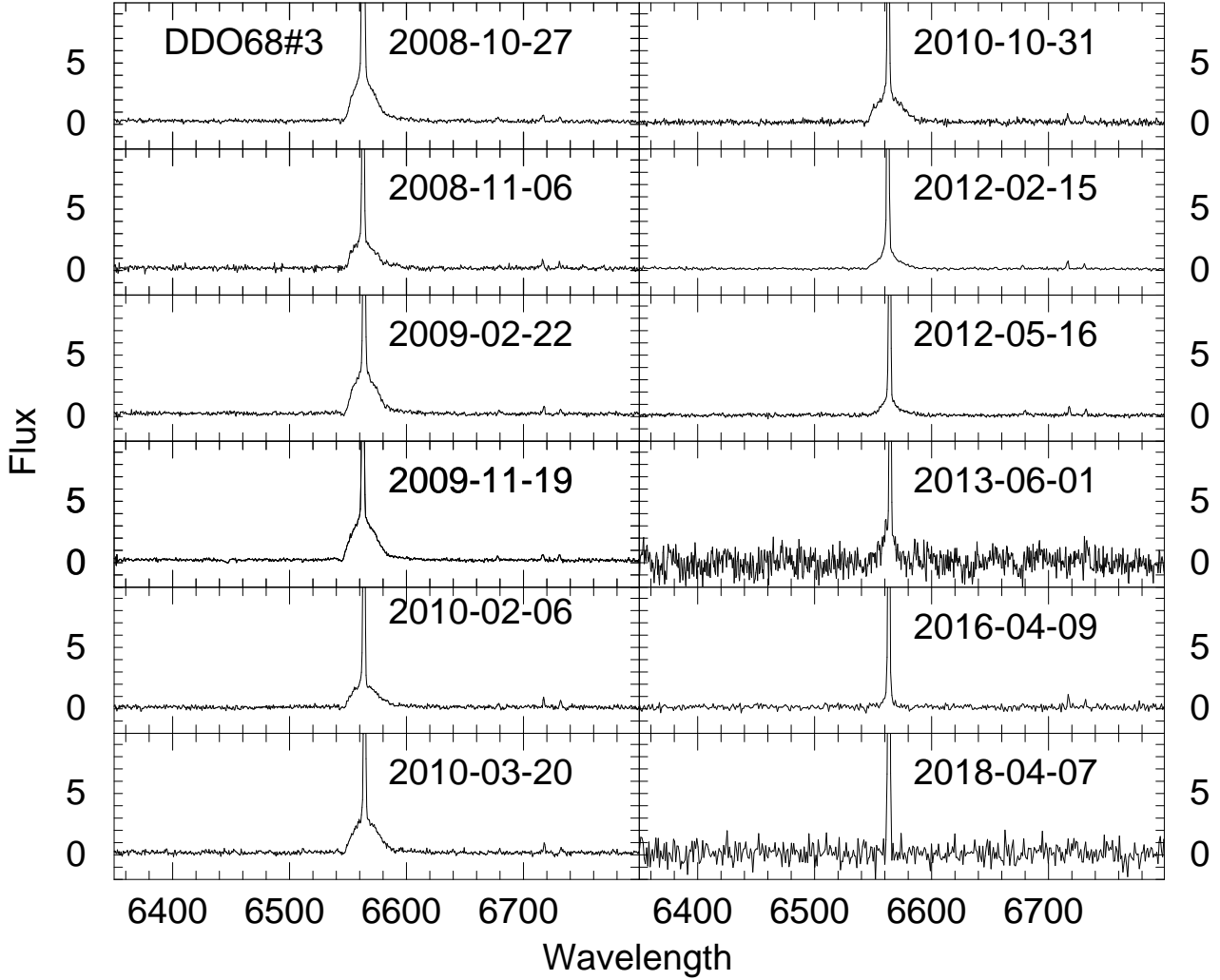}
}
\caption{The rest-frame H$\alpha$ emission line profile in the DDO\,68~\#3
spectrum at different epochs, observed with the 3.5m APO telescope. Wavelengths
are in \AA\ and fluxes are in units of 10$^{-16}$ erg s$^{-1}$ cm$^{-2}$ \AA$^{-1}$.}
\label{fig6}
\end{figure*}

\begin{figure*}
\hbox{
\includegraphics[angle=-90,width=0.48\linewidth]{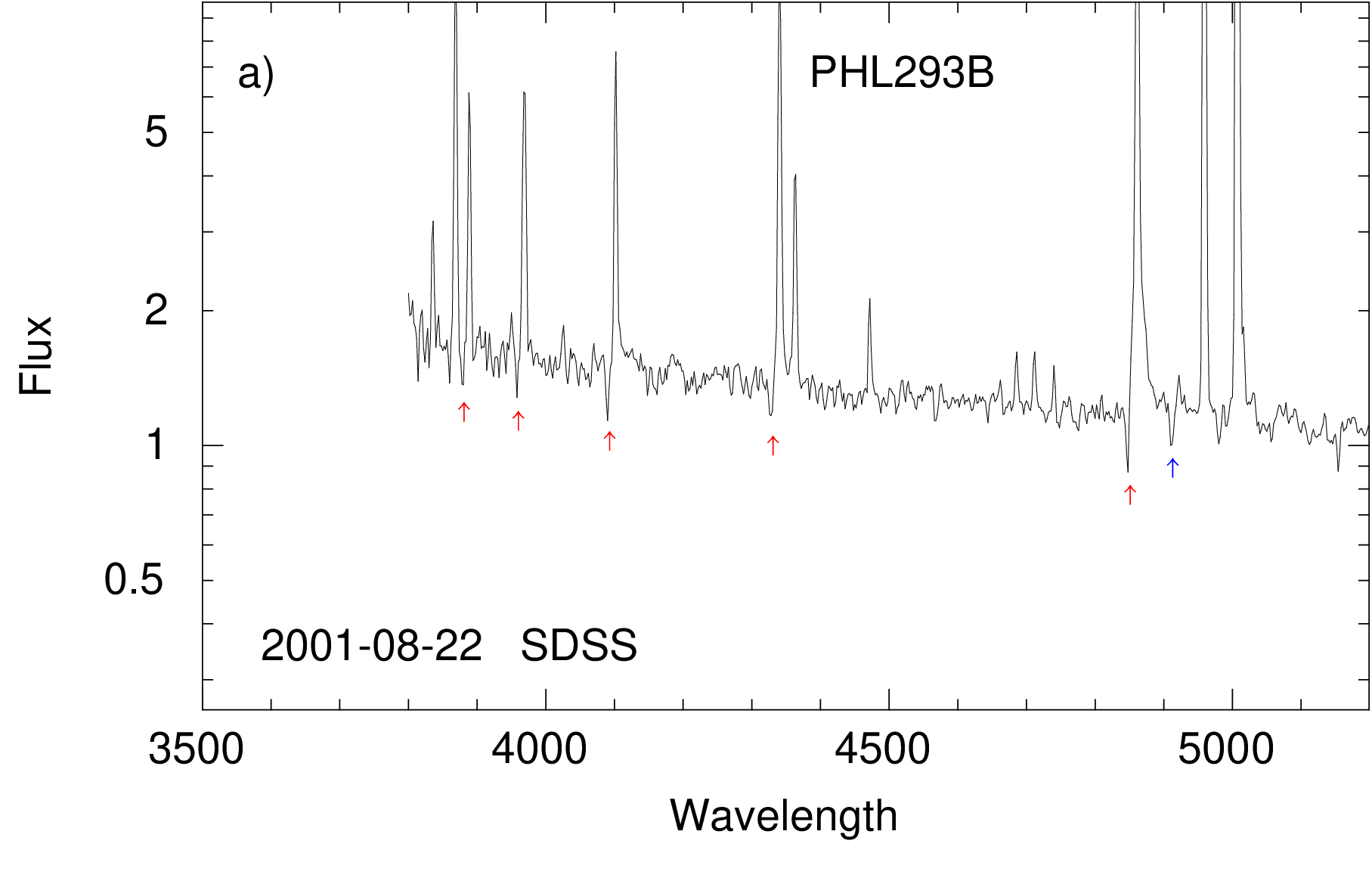}
\includegraphics[angle=-90,width=0.48\linewidth]{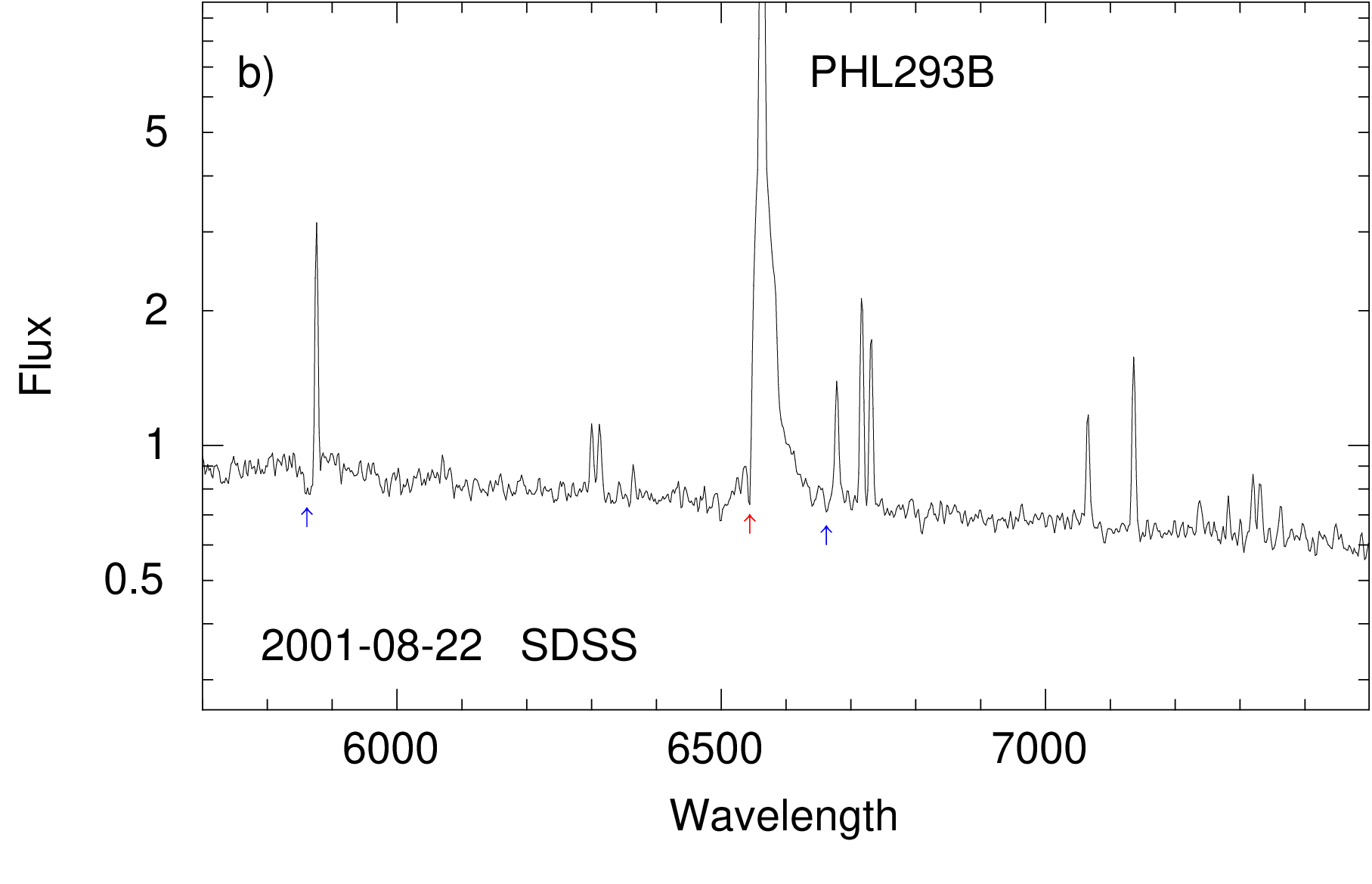}
}
\hbox{
\includegraphics[angle=-90,width=0.48\linewidth]{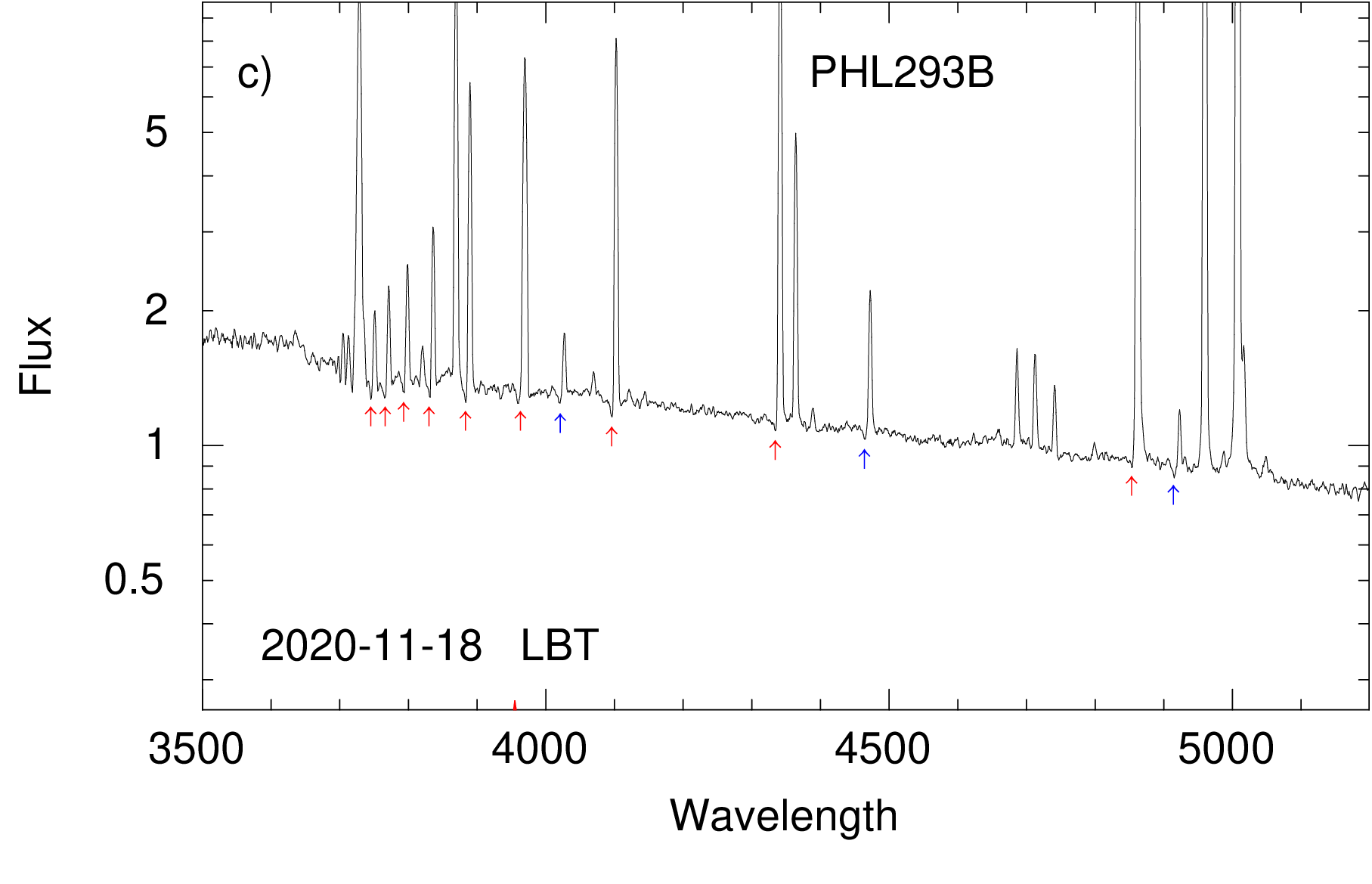}
\includegraphics[angle=-90,width=0.48\linewidth]{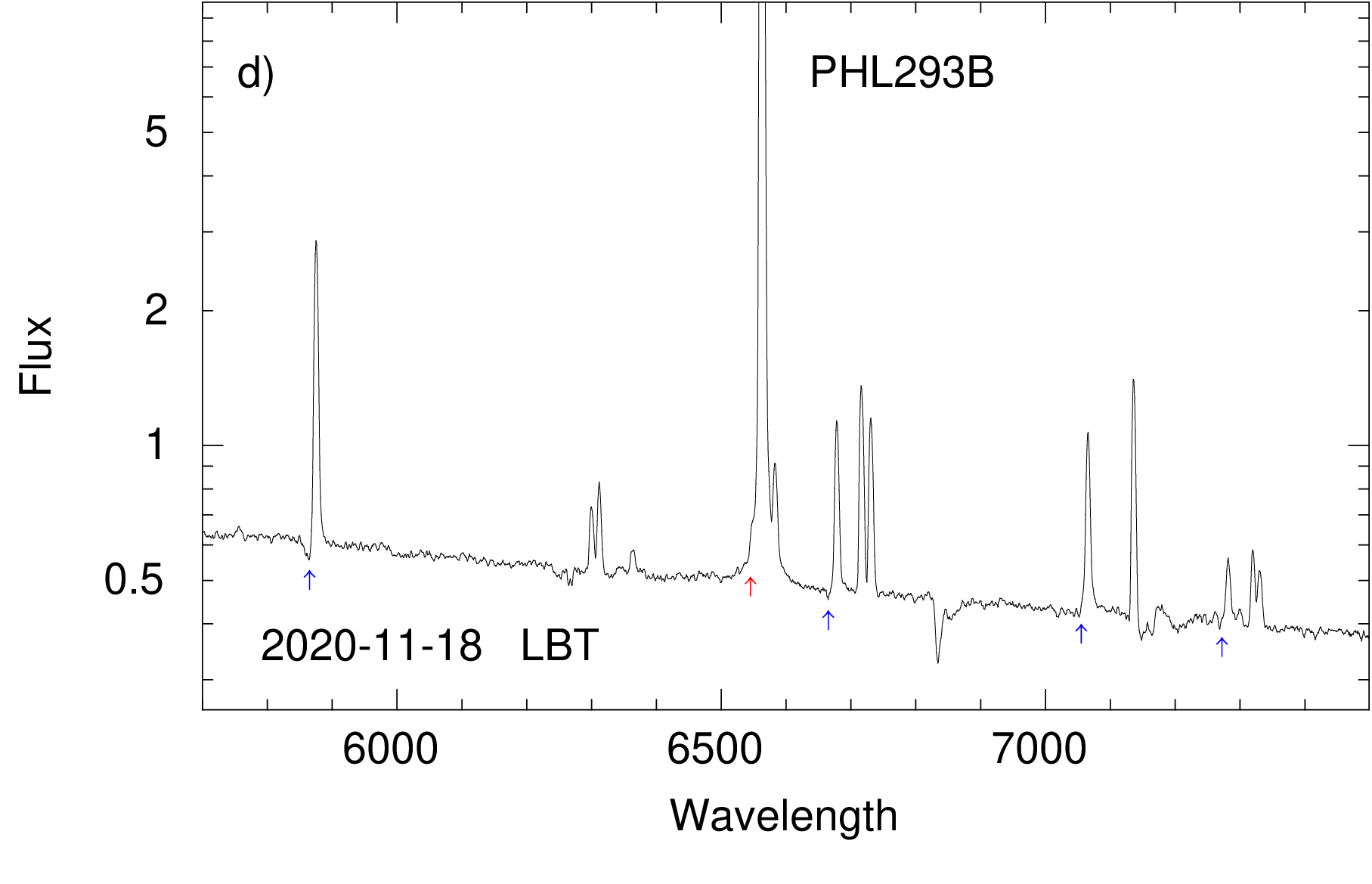}
}
\caption{The most time-separated spectra of PHL\,293B. 
Panels a) and b) show blue and red parts of the SDSS spectrum obtained  
 with the 2.5m APO telescope on August 22, 2001. 
Panels c) and d) show the spectrum
  obtained with LBT/MODS on November 18, 2018.
  Absorption features of P Cygni profiles for hydrogen and helium emission
  lines are indicated in the rest-frame spectra by red and blue arrows,
  respectively. Wavelengths are in \AA\ and fluxes are in 
  units of 10$^{-16}$ erg s$^{-1}$ cm$^{-2}$ \AA$^{-1}$.}
\label{fig6a}
\end{figure*}

\begin{figure*}
\hbox{
\includegraphics[angle=-90,width=0.98\linewidth]{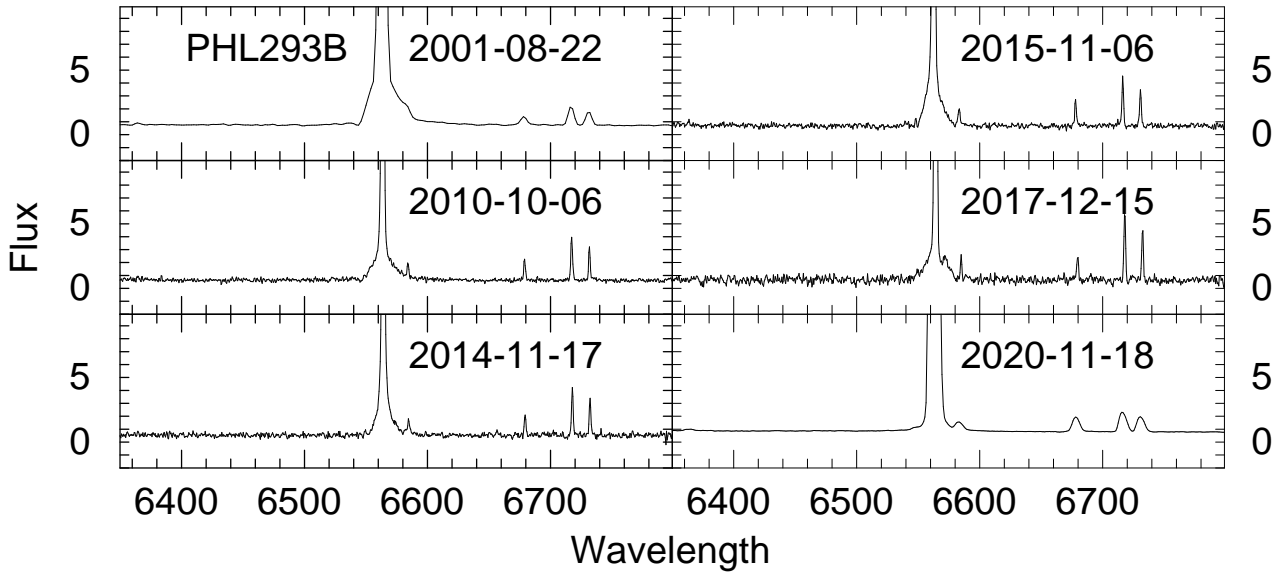}
}
\caption{The rest-frame H$\alpha$ emission line profiles in the PHL\,293B
spectra at the different epochs listed in Table~\ref{tab7-PHL-G}. Wavelengths are
in \AA\ and fluxes are in units of 10$^{-16}$ erg s$^{-1}$ cm$^{-2}$ \AA$^{-1}$.}
\label{fig7}
\end{figure*}

\begin{figure}
\hbox{
\includegraphics[angle=-90,width=0.49\linewidth]{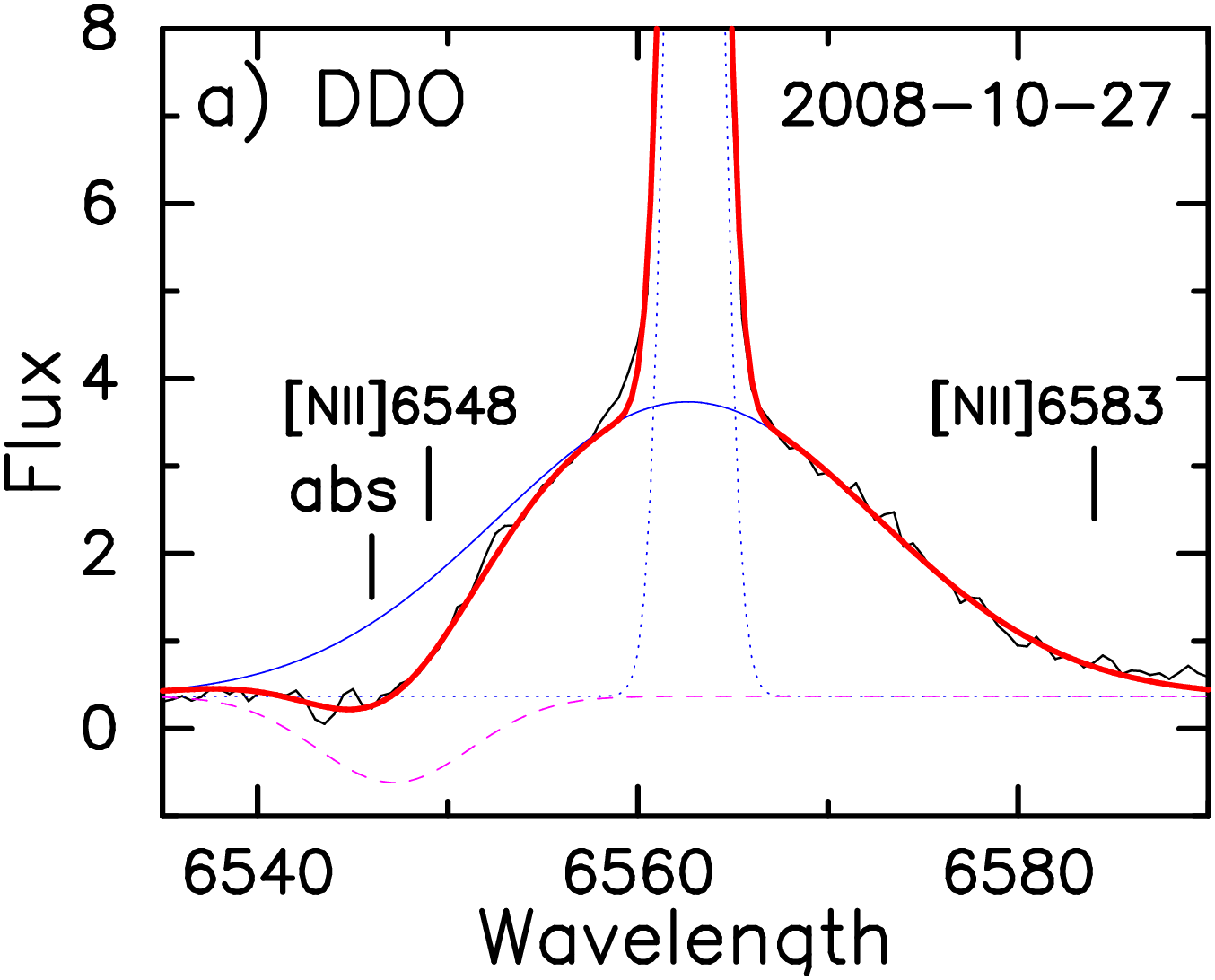}
\includegraphics[angle=-90,width=0.49\linewidth]{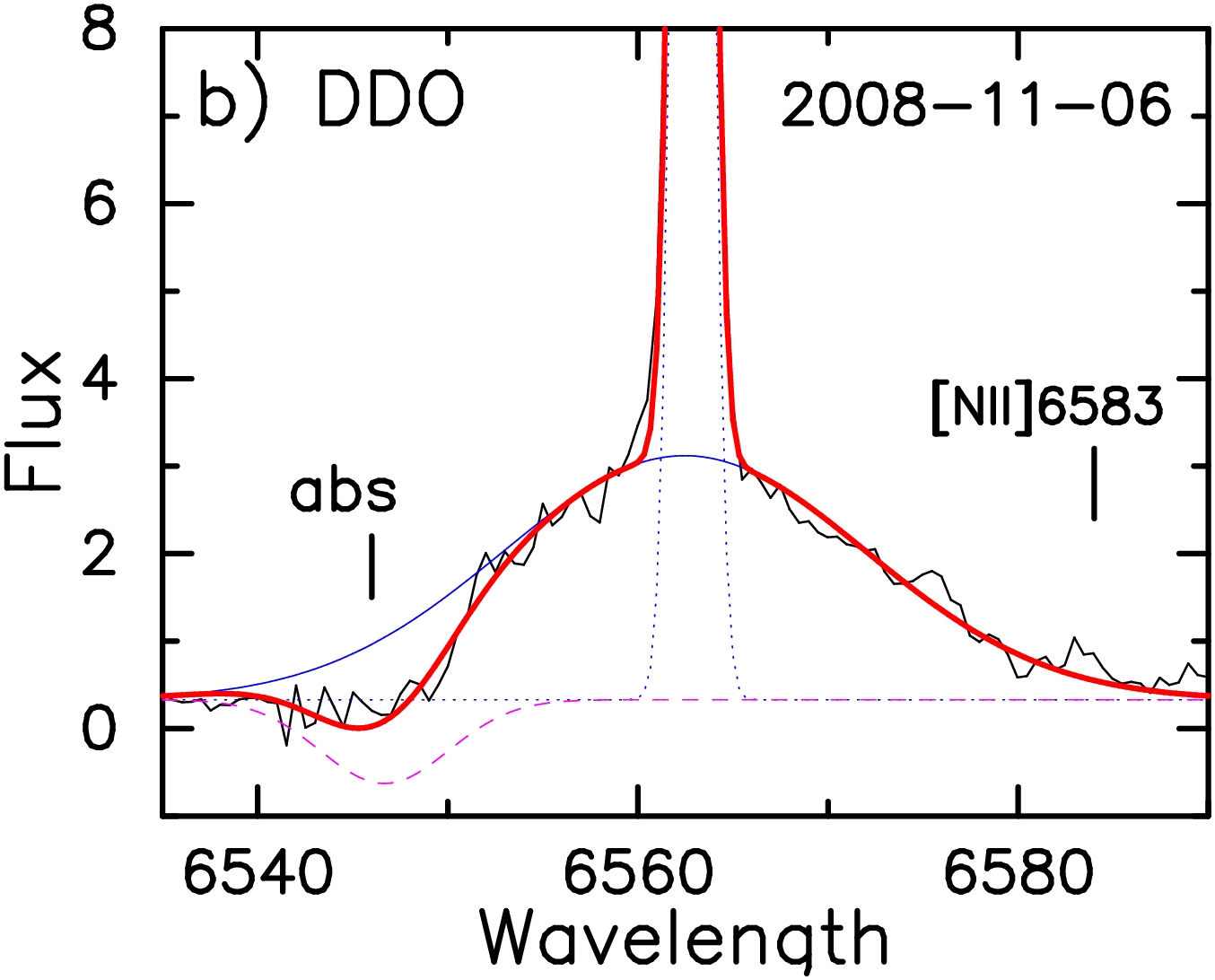}
}
\hbox{
\includegraphics[angle=-90,width=0.49\linewidth]{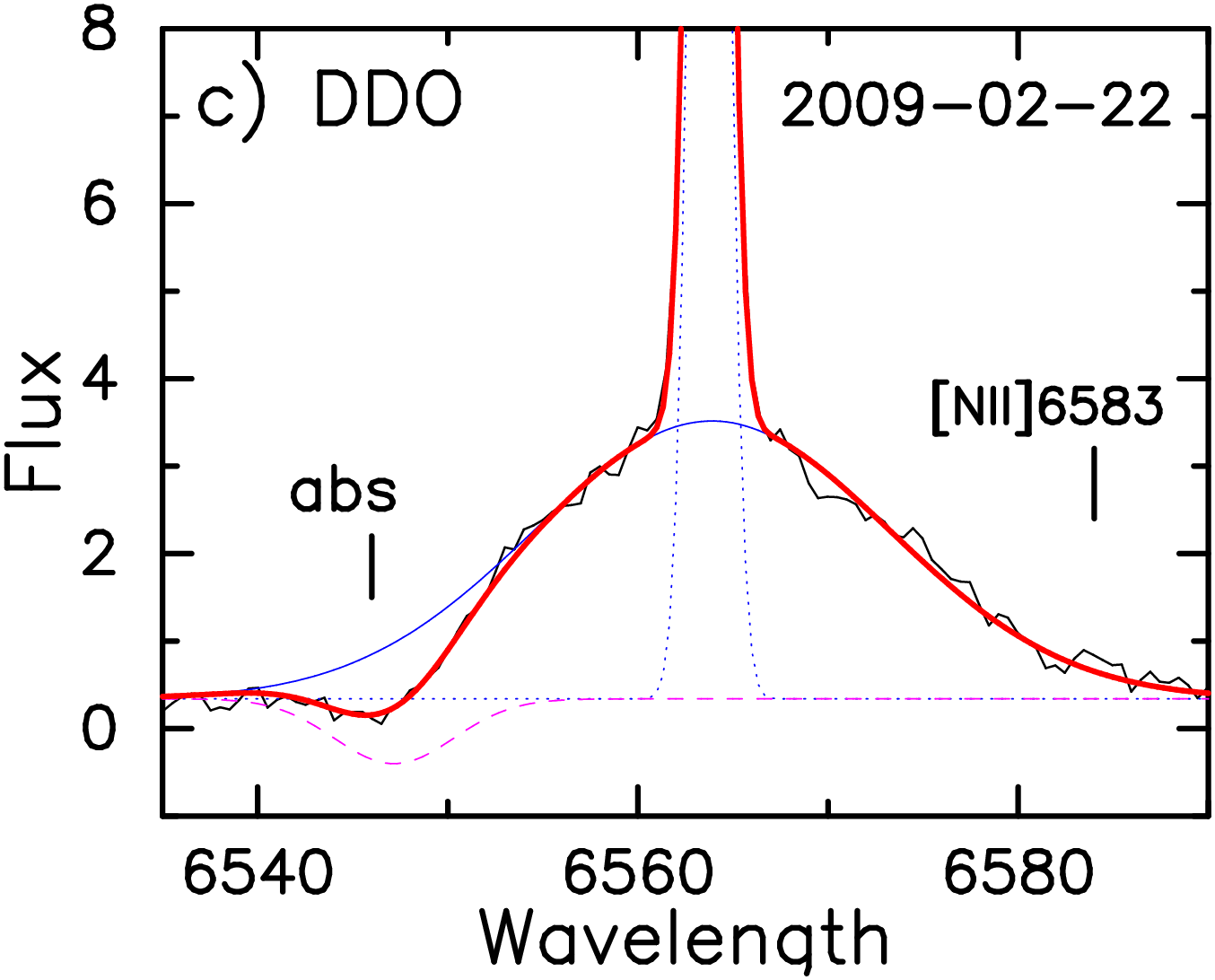}
\includegraphics[angle=-90,width=0.49\linewidth]{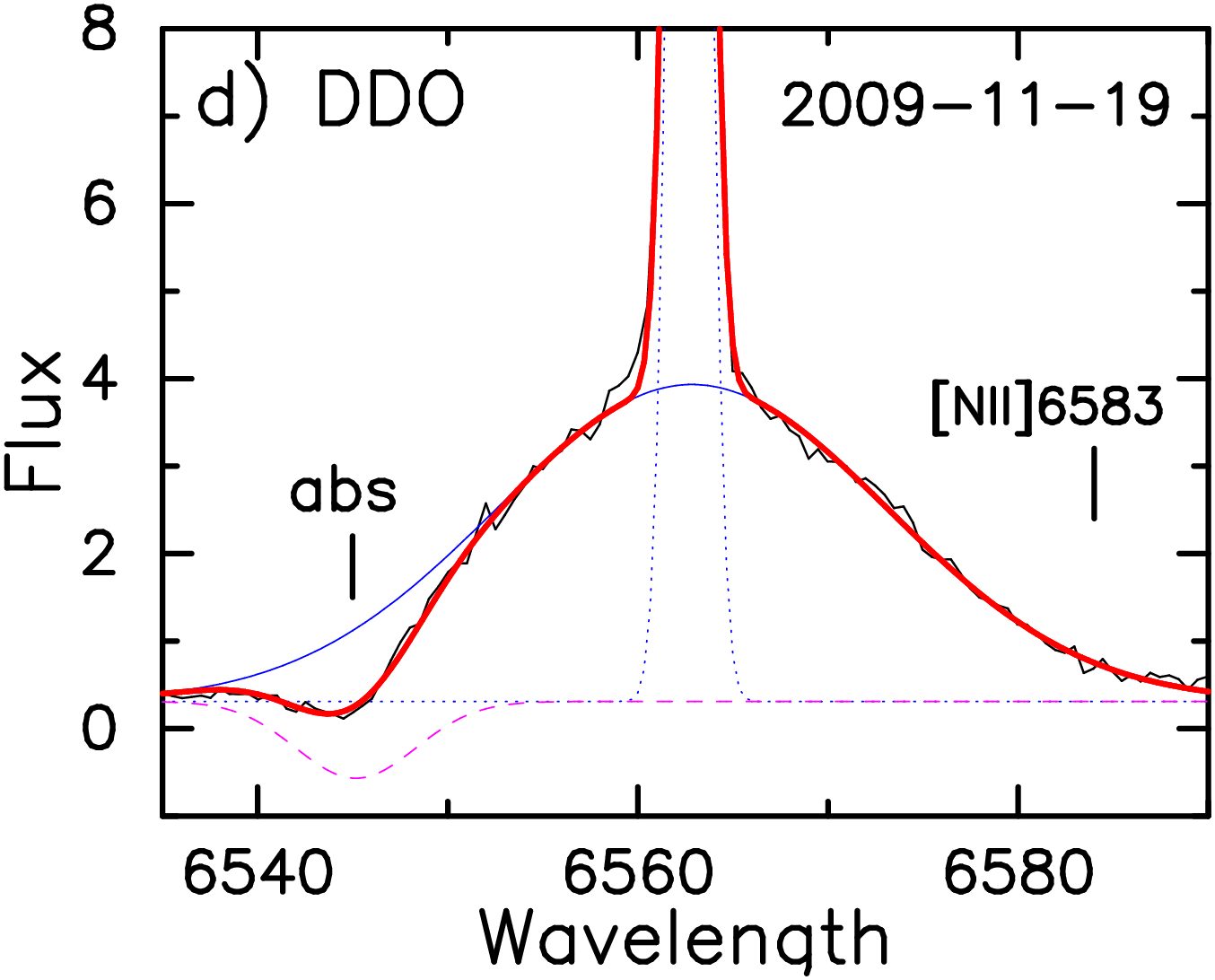}
}
\hbox{
\includegraphics[angle=-90,width=0.49\linewidth]{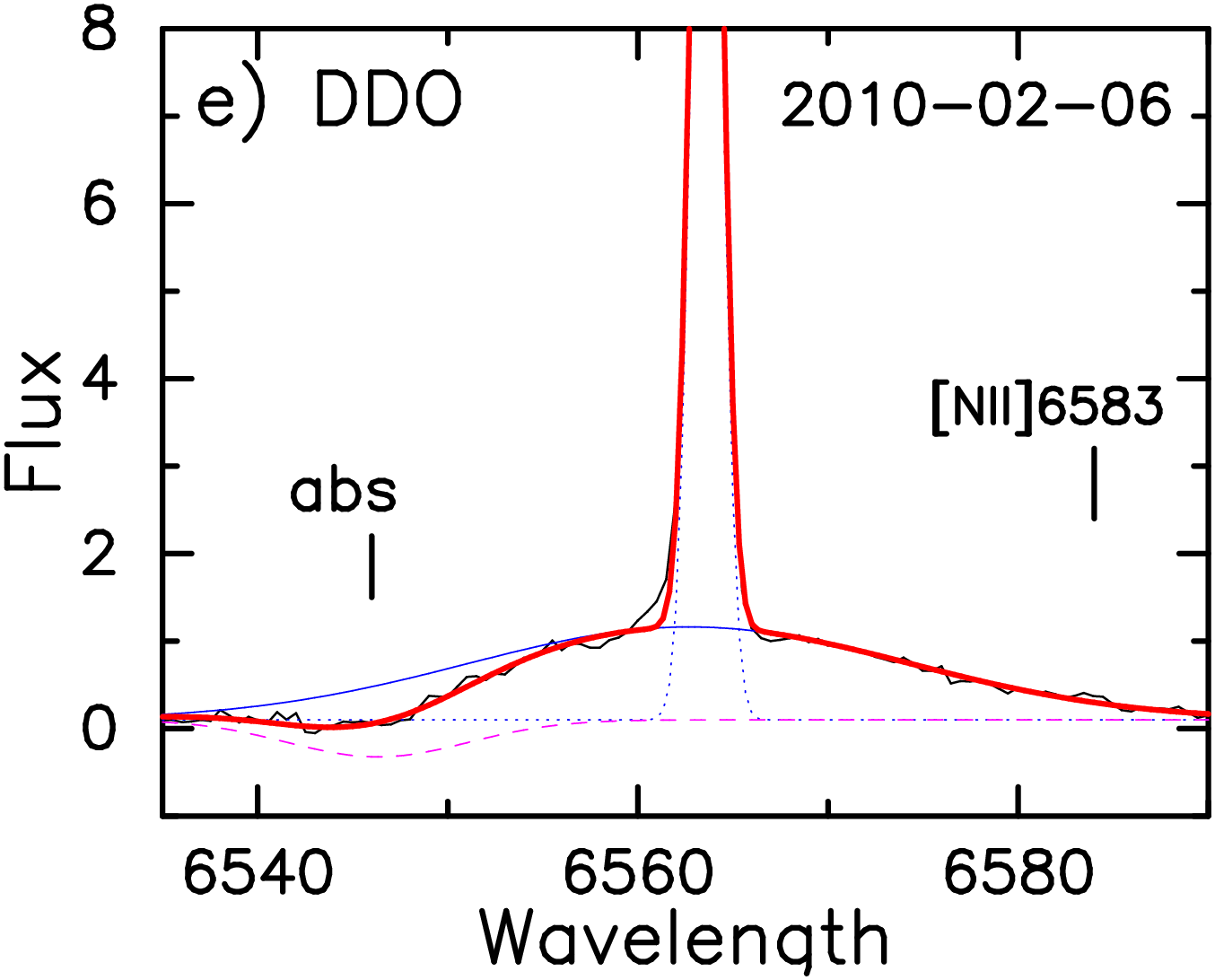}
\includegraphics[angle=-90,width=0.49\linewidth]{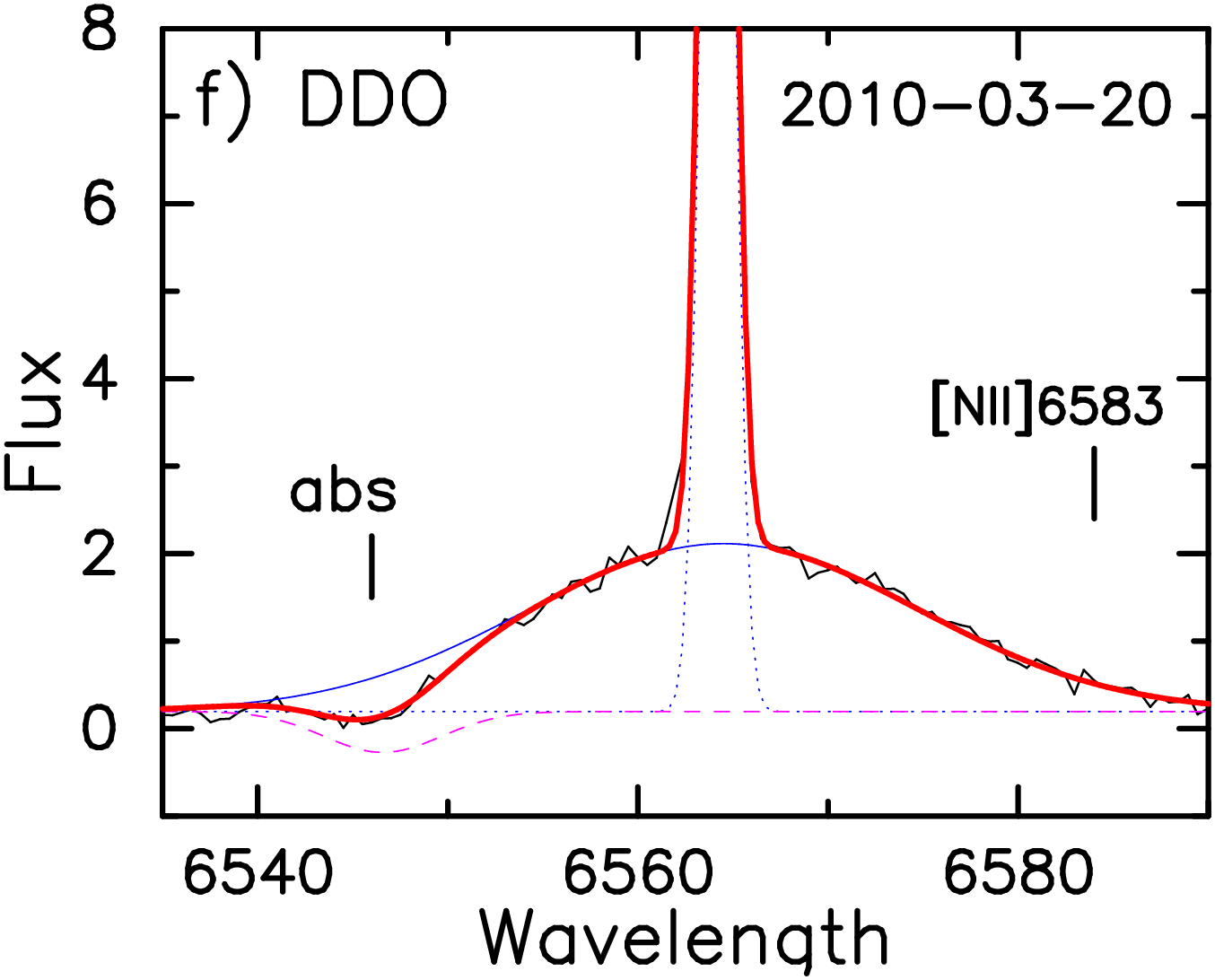}
}
\hbox{
\includegraphics[angle=-90,width=0.49\linewidth]{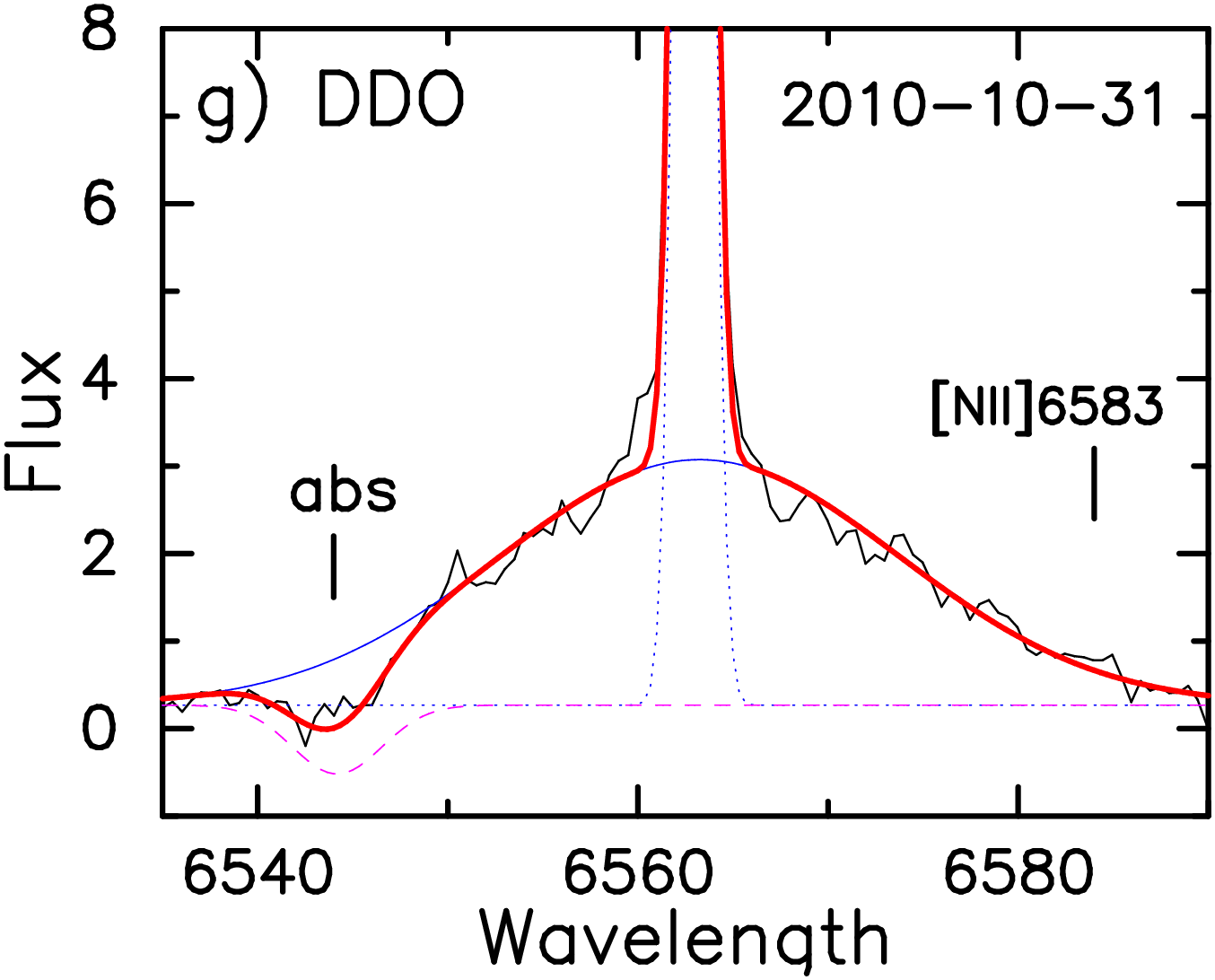}
\includegraphics[angle=-90,width=0.49\linewidth]{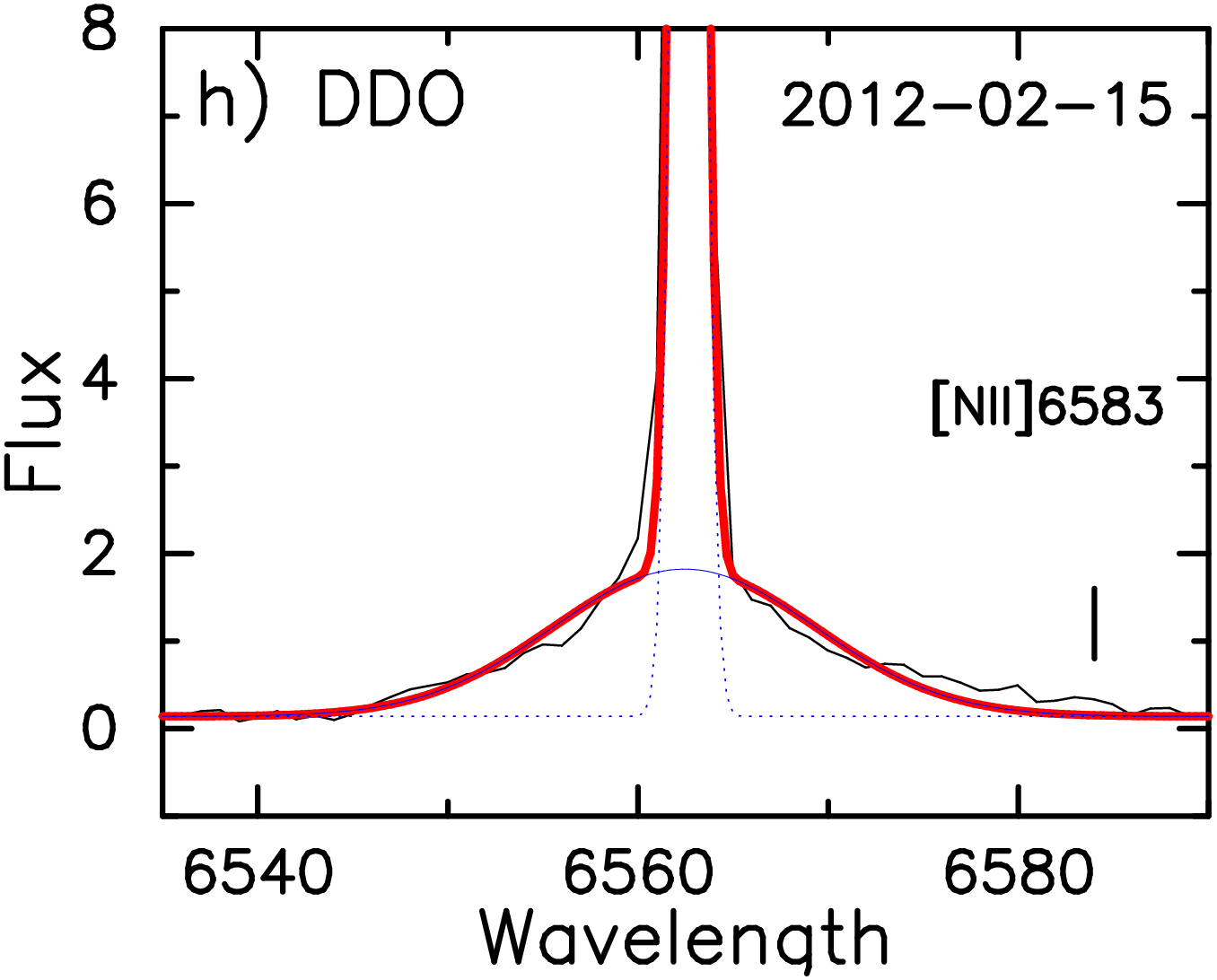}
}
\hbox{
\includegraphics[angle=-90,width=0.49\linewidth]{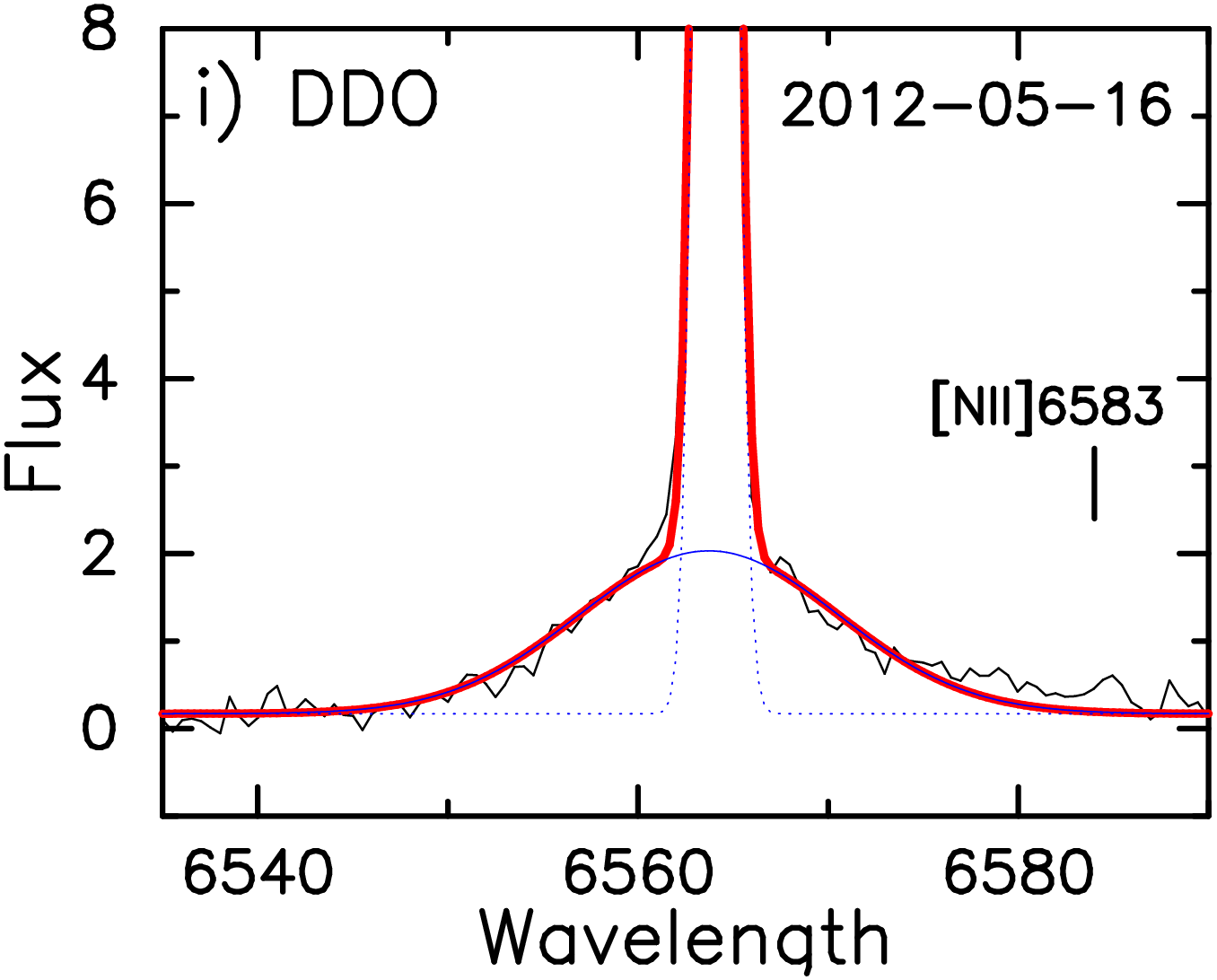}
\includegraphics[angle=-90,width=0.49\linewidth]{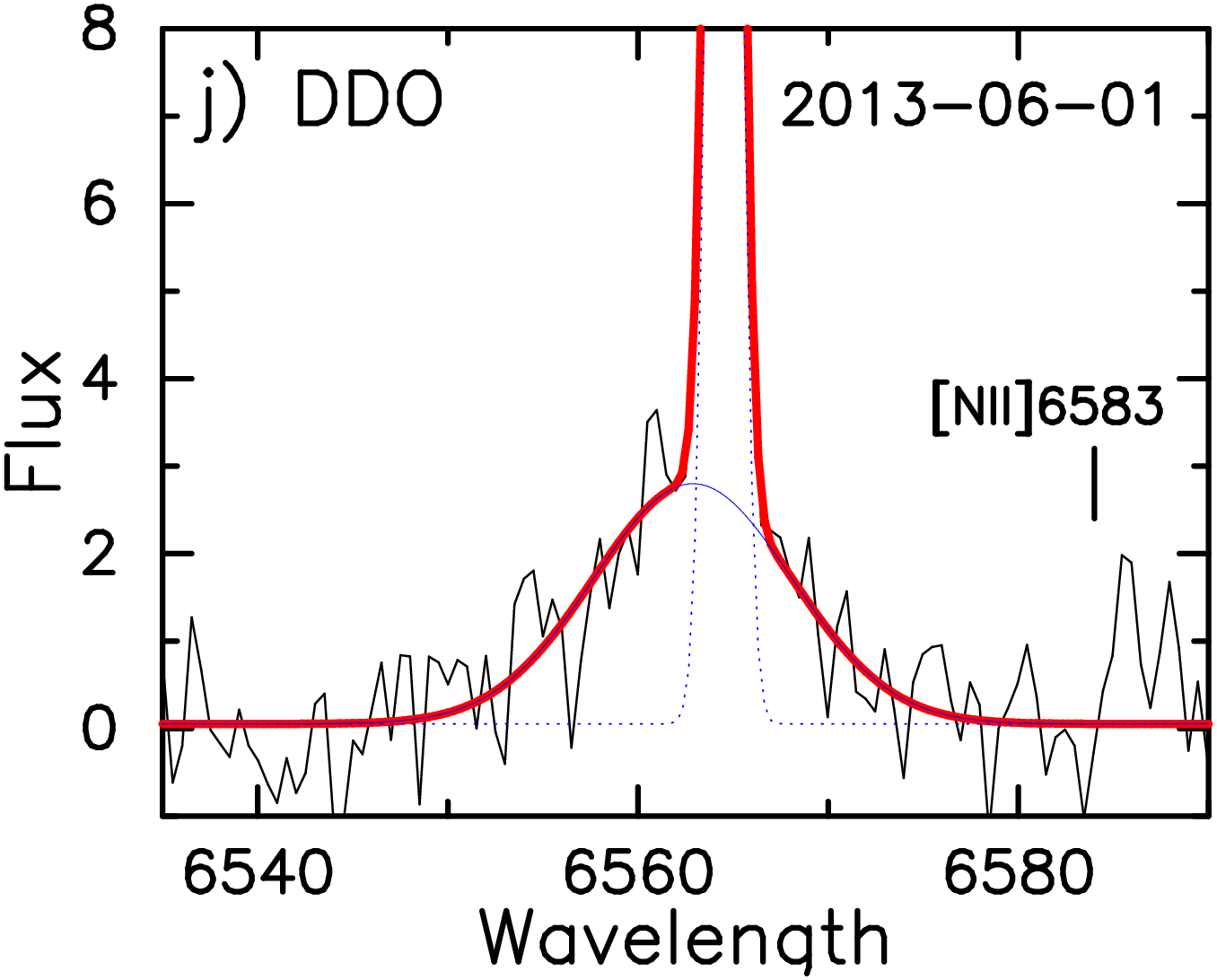}
}
\hbox{
\includegraphics[angle=-90,width=0.49\linewidth]{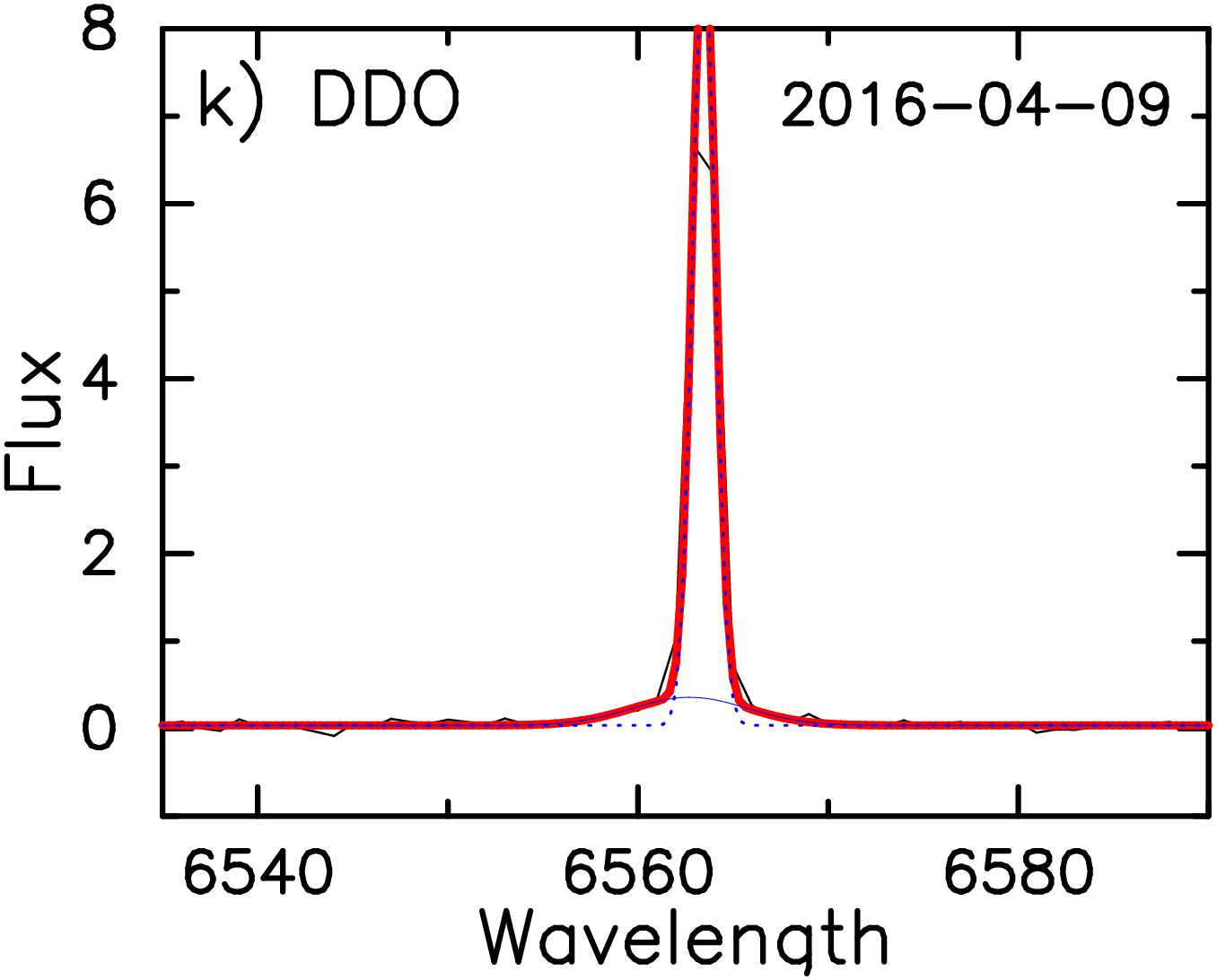}
\includegraphics[angle=-90,width=0.49\linewidth]{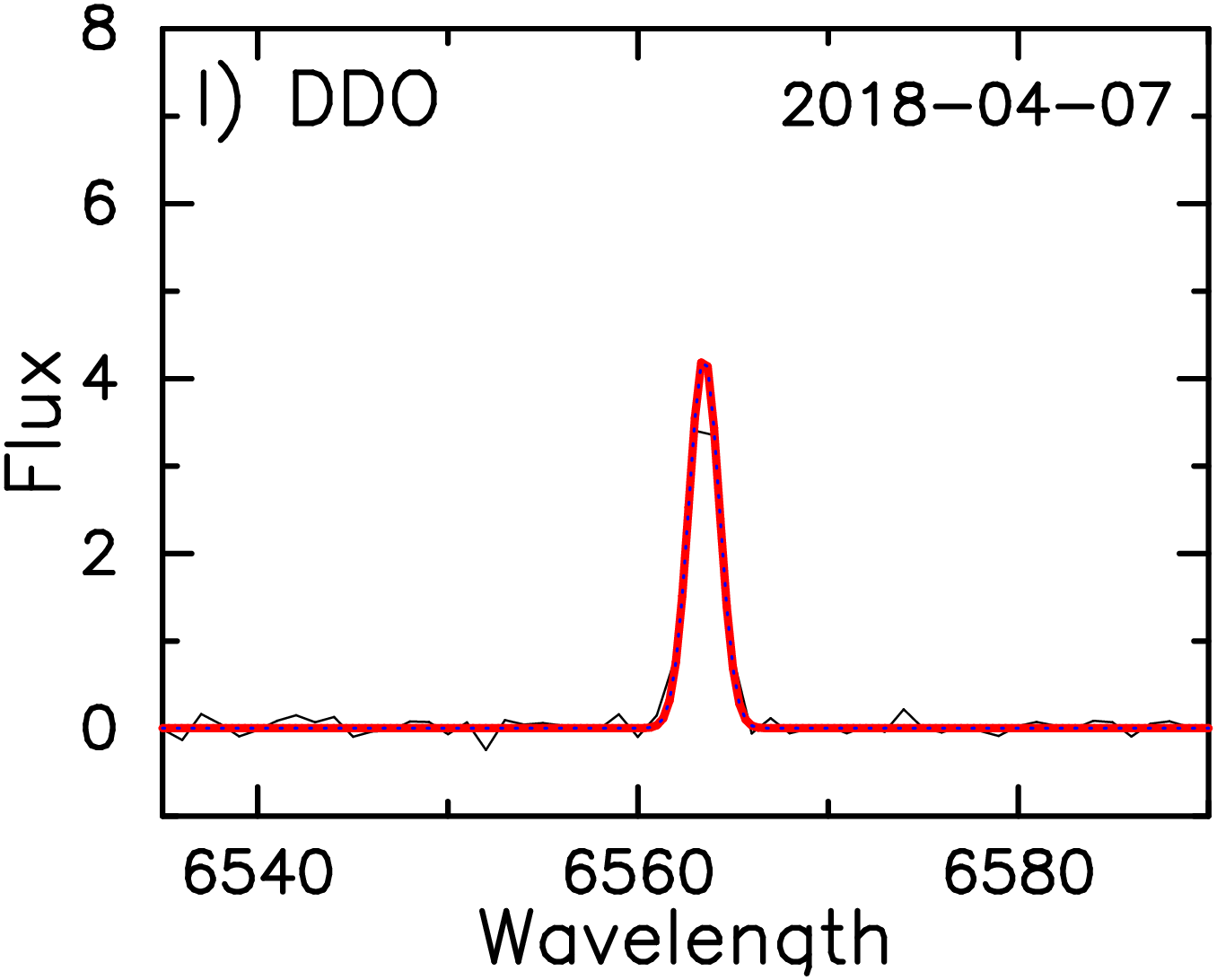}
}
\caption{Decomposition by Gaussians of the H$\alpha$ narrow (blue dotted lines) and
broad emission (blue solid lines) profiles in the spectra of DDO\,68~\#3, listed 
in Table~\ref{tab7-DDO-G} and shown in 
Fig.~\ref{fig6}. The observed profile is shown by the black solid line whereas
the summed flux of narrow+broad components is represented by the red solid line.
Wavelengths are in \AA\ and fluxes are in units of
10$^{-16}$ erg s$^{-1}$ cm$^{-2}$ \AA$^{-1}$.}
\label{fig2}
\end{figure}

\begin{figure}
\hbox{
\includegraphics[angle=-90,width=0.49\linewidth]{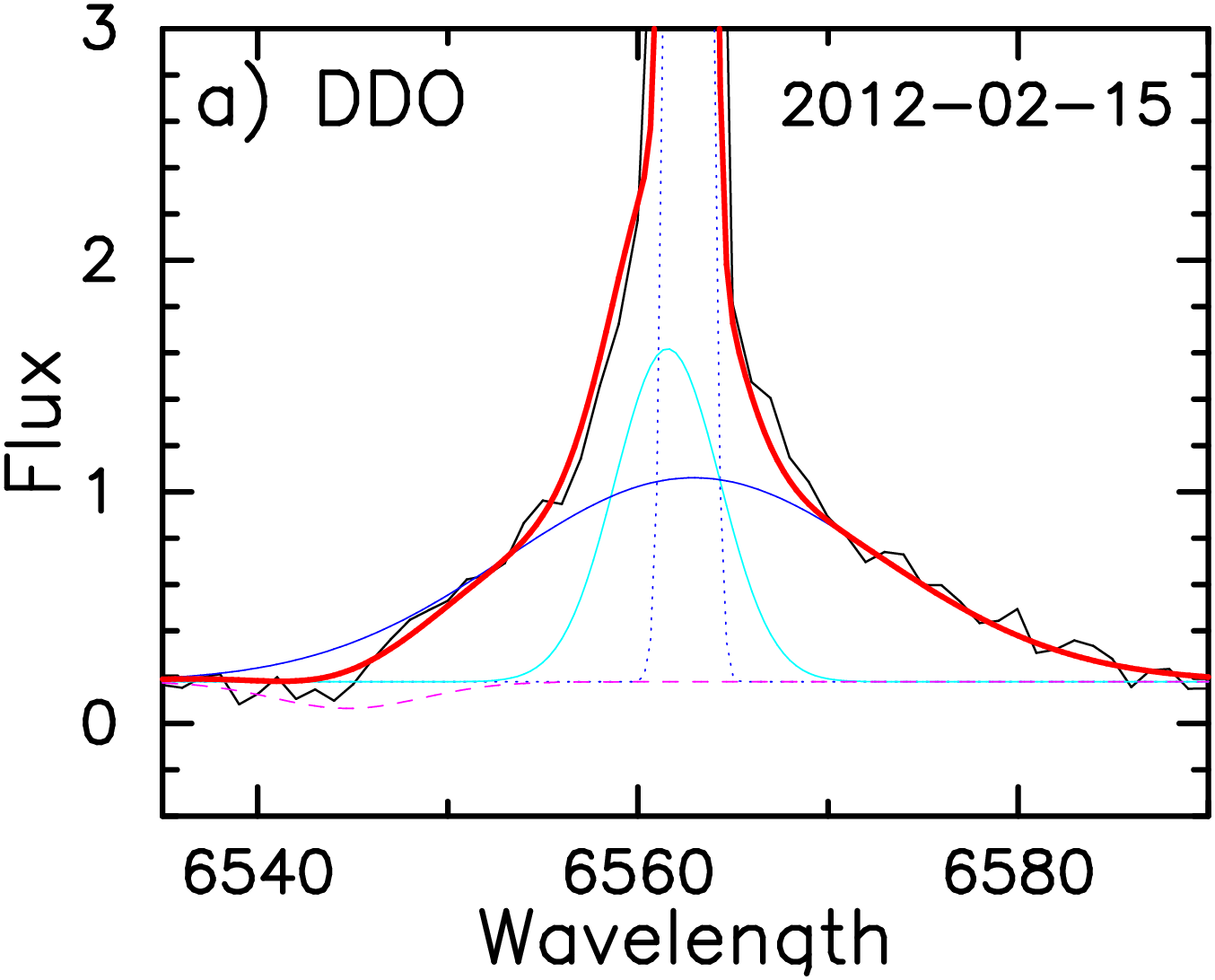}
\includegraphics[angle=-90,width=0.49\linewidth]{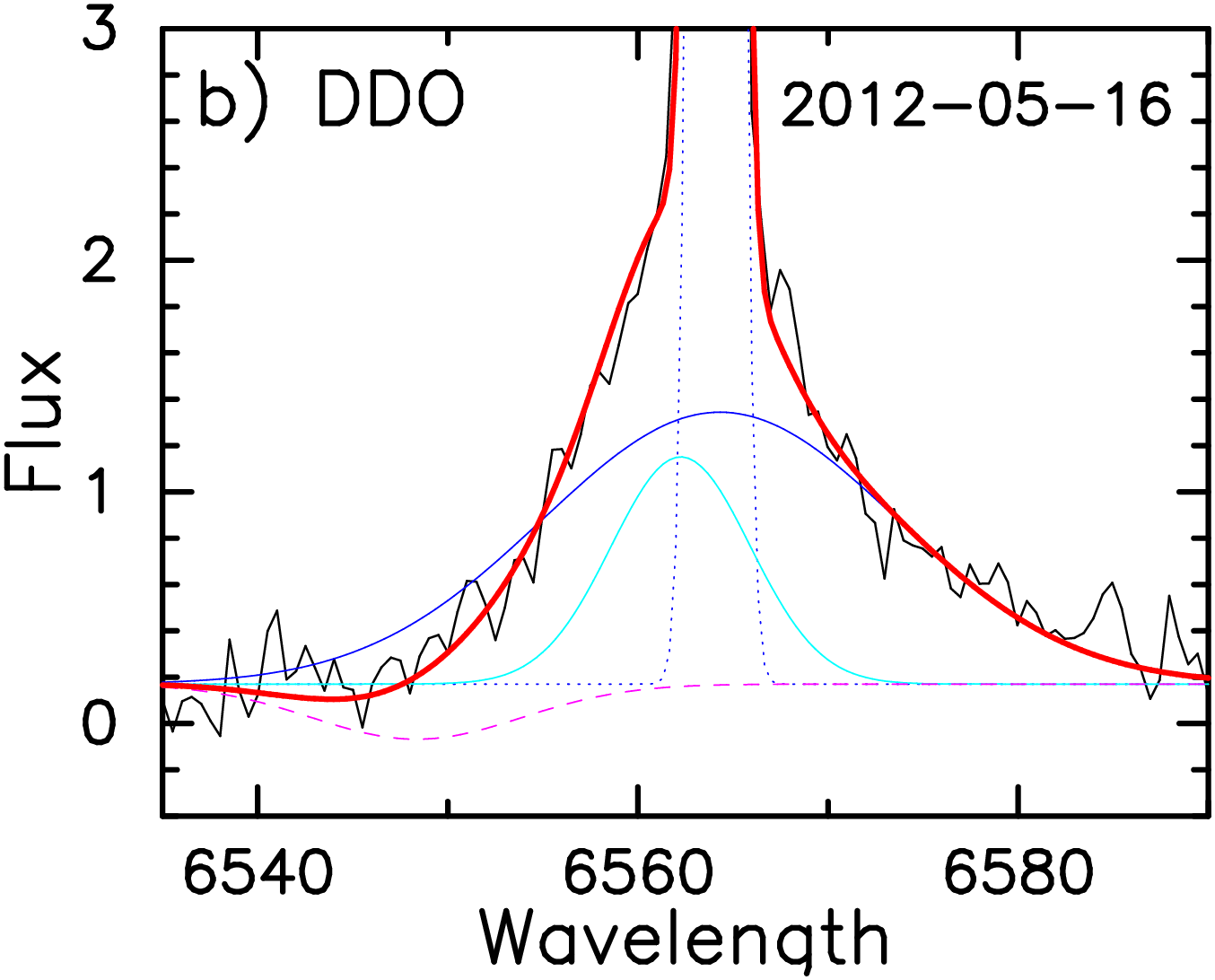}
}
\caption{The observed profile of H$\alpha$ in DDO\,68~\#3 with an excess flux at high velocities in
redward wings. This excess is seen only for the observations on 15 February and 16 May 2012
(see Fig.~\ref{fig2}h and  Fig.~\ref{fig2}i).
Three Gaussians were used to fit the narrow, broad and very broad components (blue dots, turquoise solid and blue solid lines, respectively). 
The total fitted profile is represented by a red solid line.  Wavelengths are
in \AA\ and fluxes are in units of 10$^{-16}$ erg s$^{-1}$ cm$^{-2}$ \AA$^{-1}$.}
\label{fig3}
\end{figure}

\begin{figure}
\hbox{
\includegraphics[angle=-90,width=0.49\linewidth]{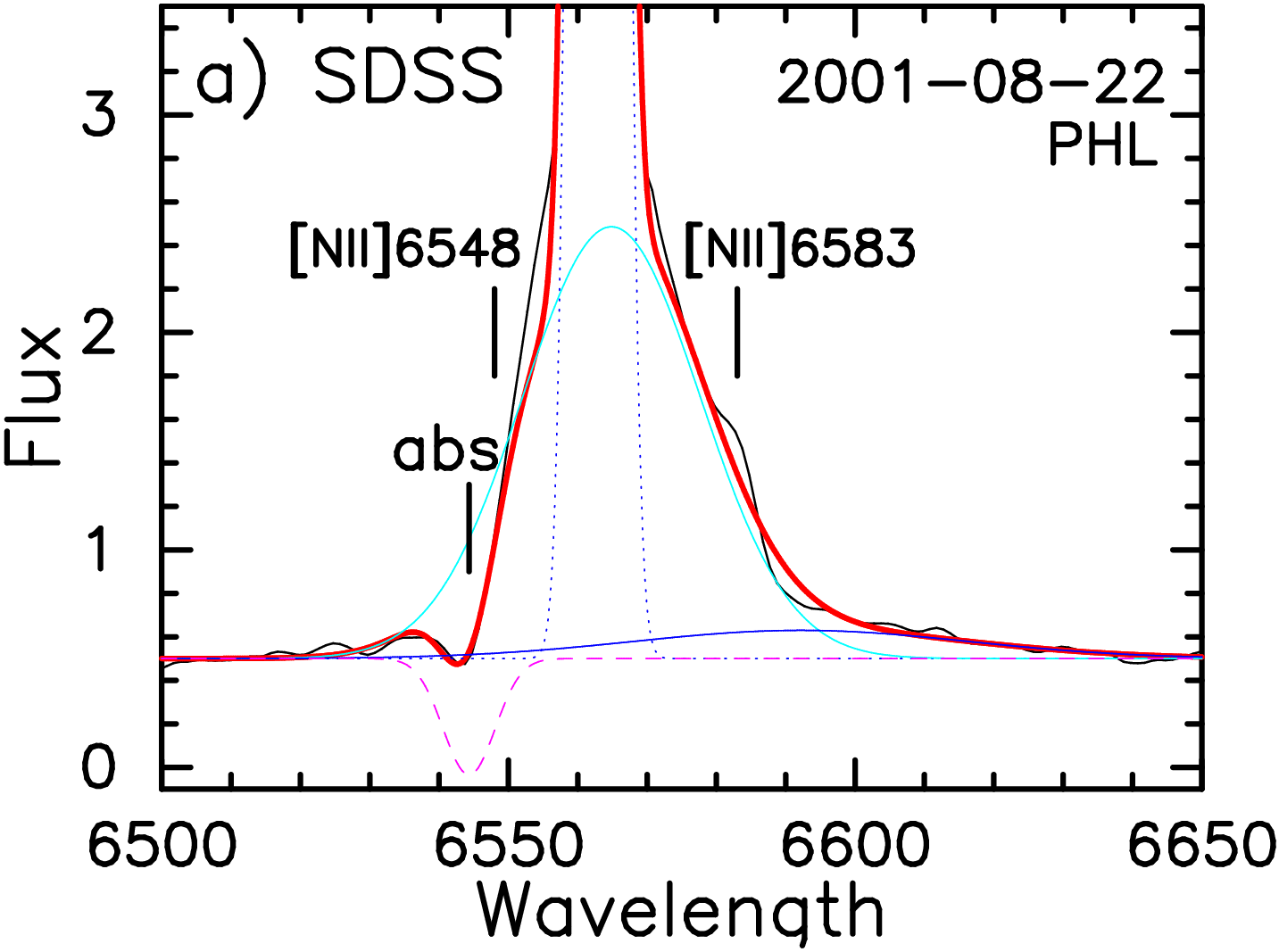}
\includegraphics[angle=-90,width=0.49\linewidth]{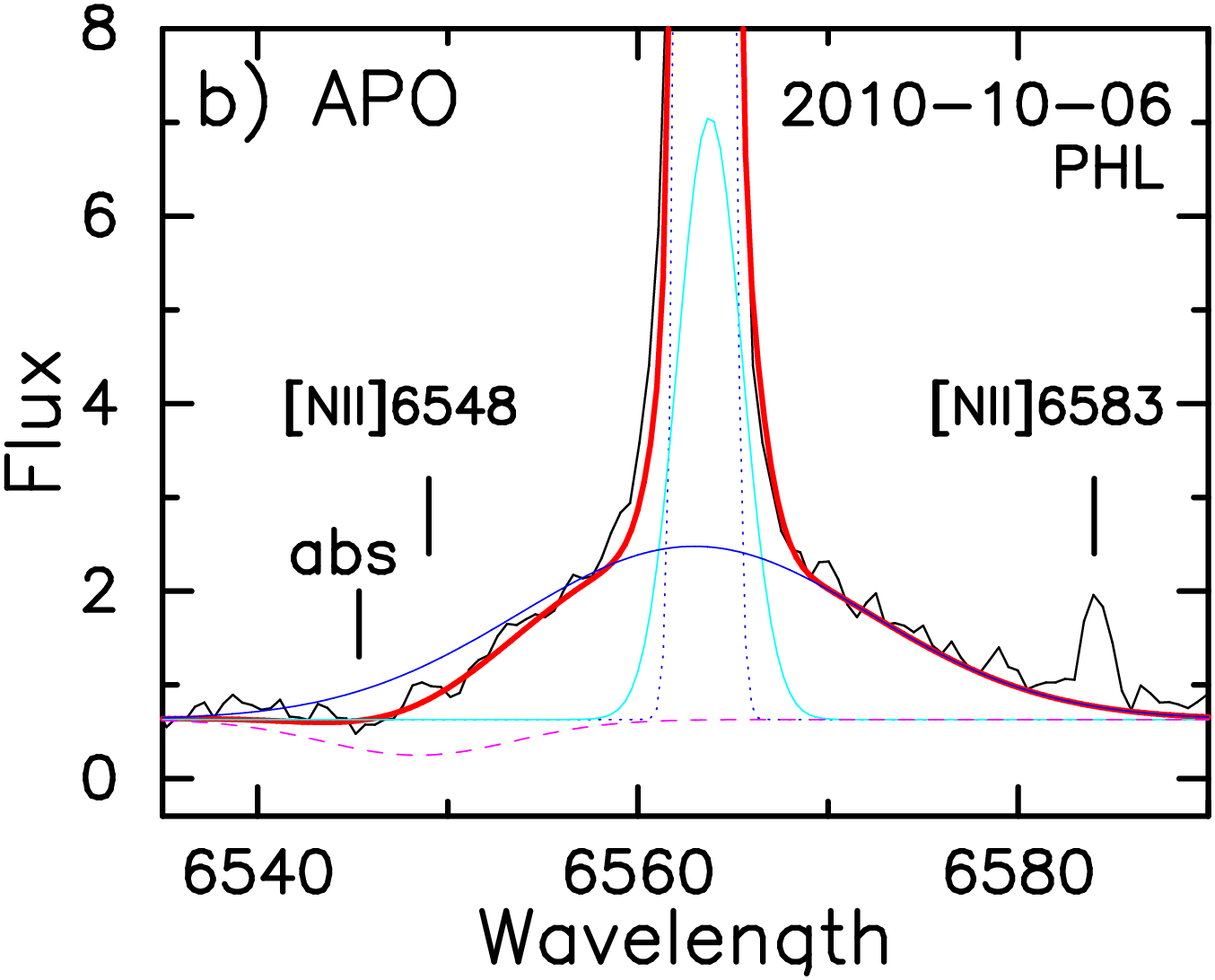}
}
\hbox{
\includegraphics[angle=-90,width=0.49\linewidth]{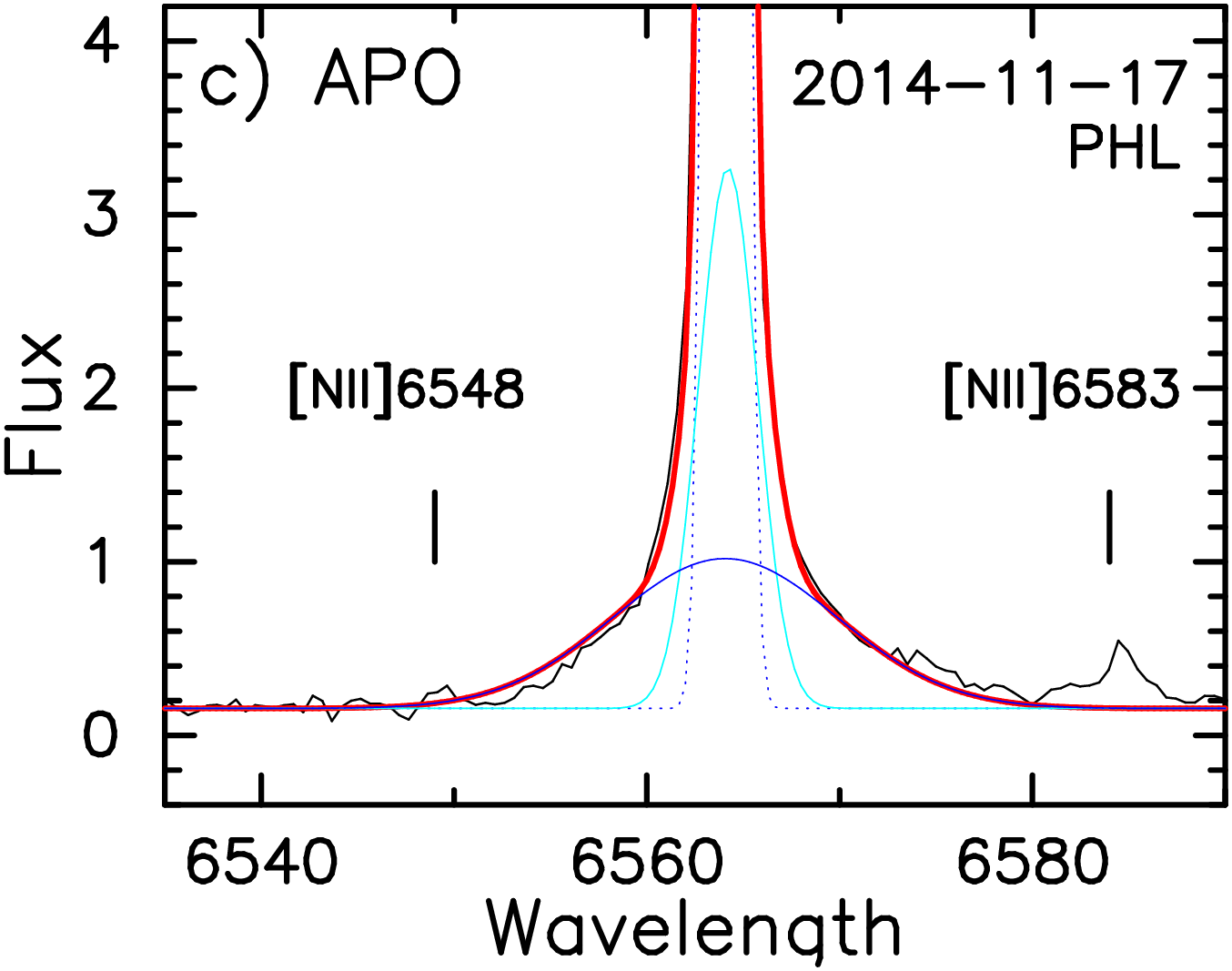}
\includegraphics[angle=-90,width=0.49\linewidth]{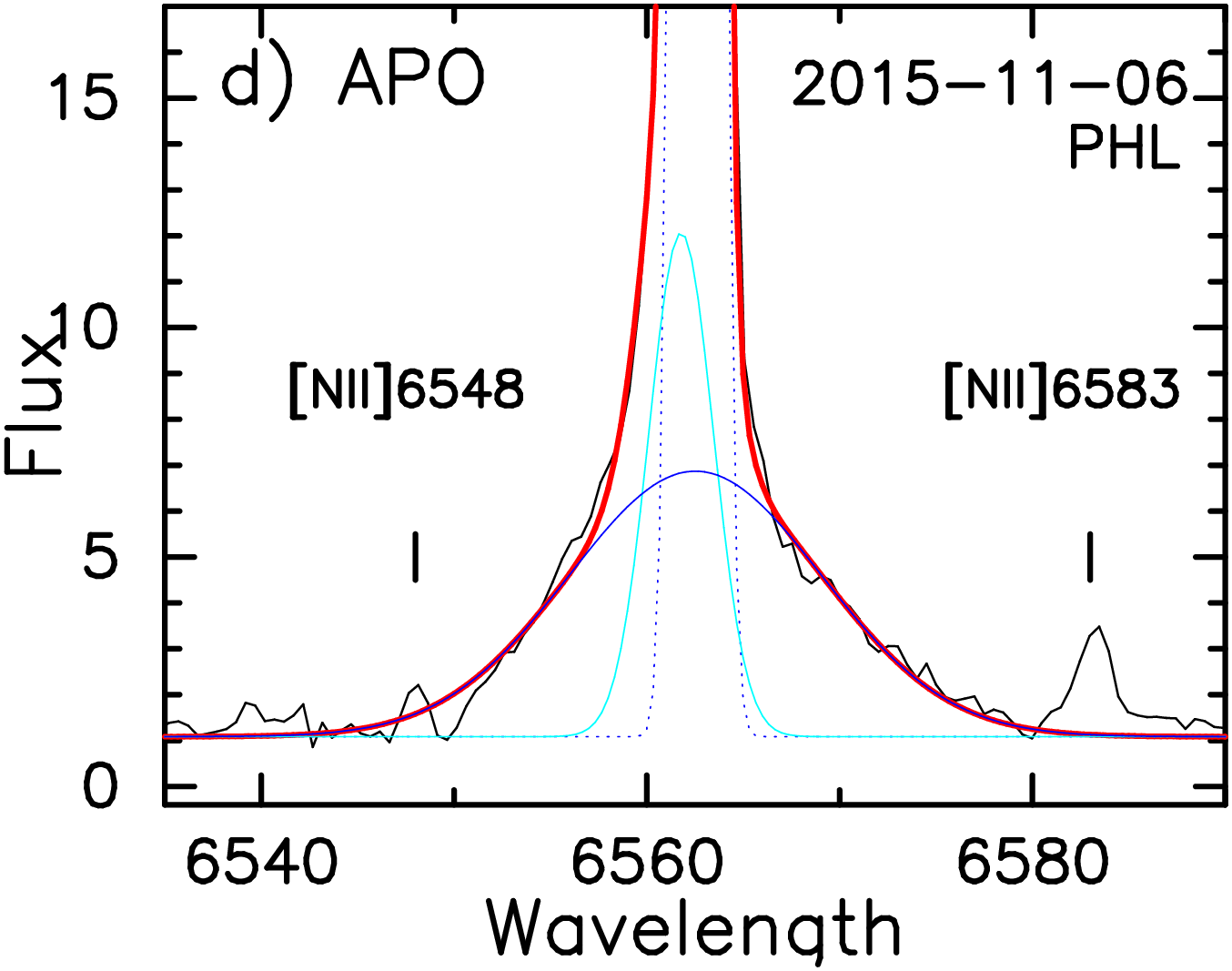}
}
\hbox{
\includegraphics[angle=-90,width=0.49\linewidth]{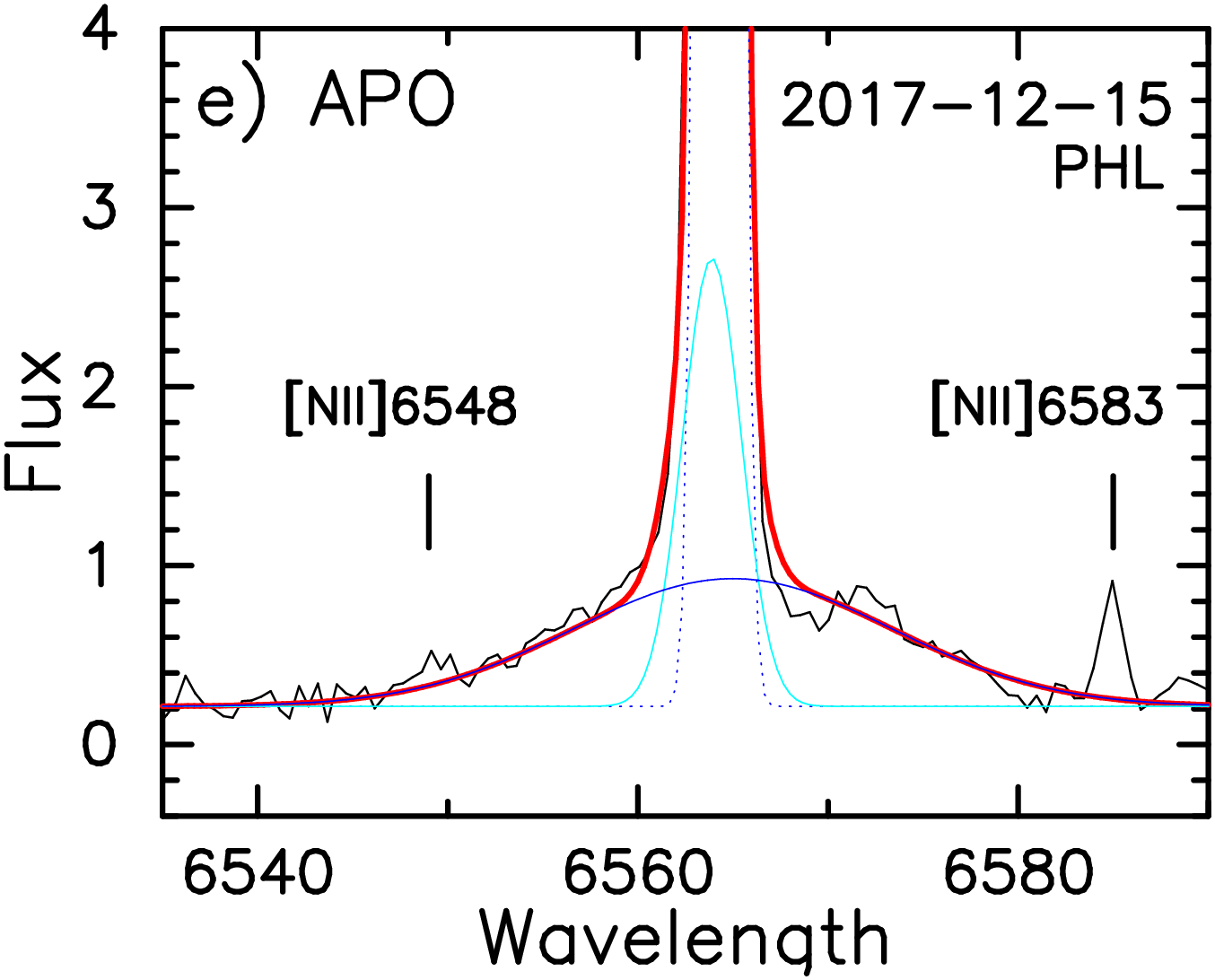}
\includegraphics[angle=-90,width=0.49\linewidth]{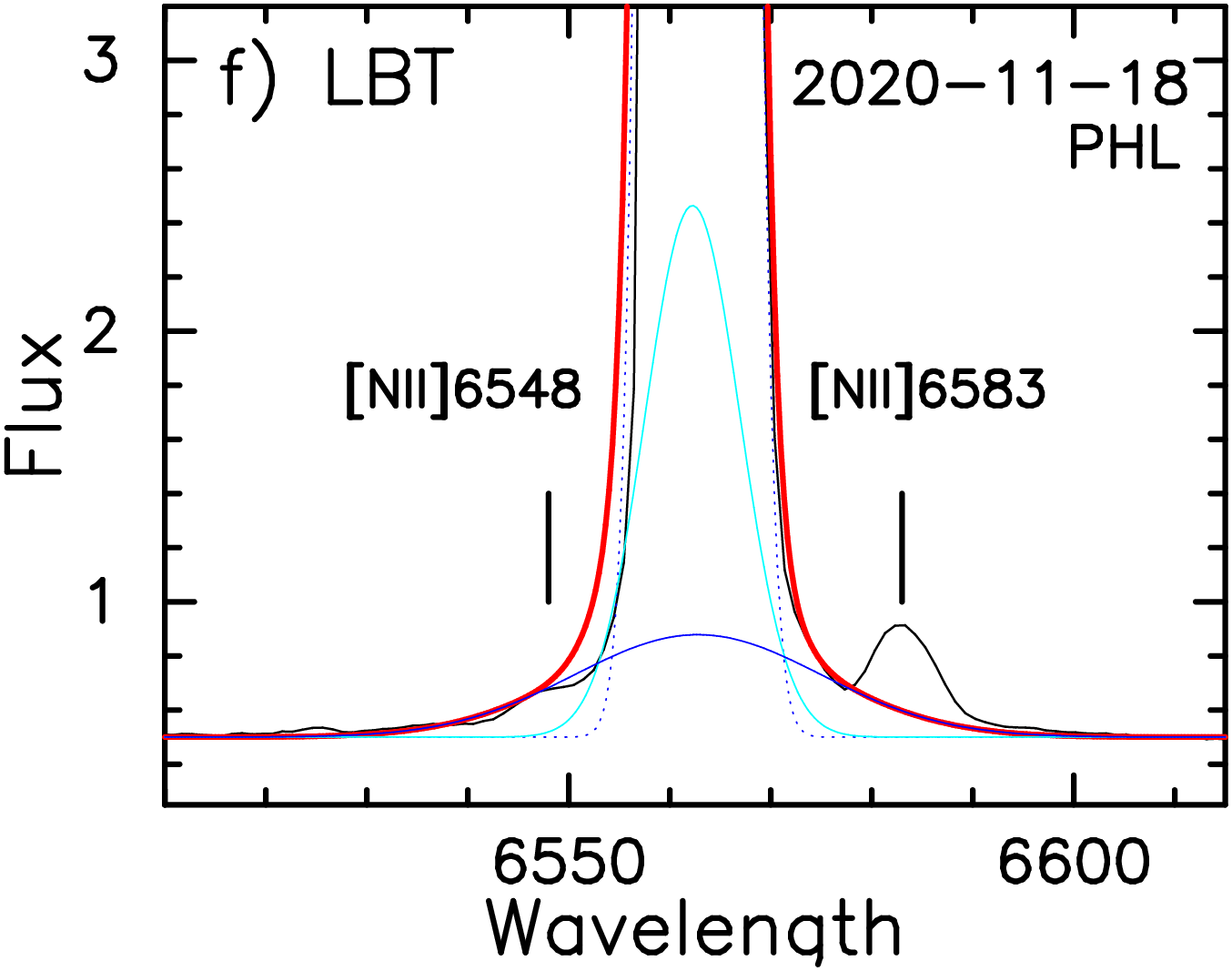}
}
\caption{The H$\alpha$ profiles in the PHL\,293B spectra obtained with
different telescopes during different epochs. The black lines represent the
observed spectra. The fits of H$\alpha$ by three or four Gaussians are shown by
blue dotted lines (narrow component), by turquoise solid lines (broad component),
by blue solid lines (very broad component) and by magenta dashed lines (for
absorption features of P Cygni profiles whenever possible). Note that, in contrast to Fig.~\ref{fig2}
and Fig.~\ref{fig3},
the y-axes have different scales, 
to better see the shape of the broad components. 
Fluxes are in units of 10$^{-16}$ erg s$^{-1}$ cm$^{-2}$ \AA$^{-1}$
and wavelengths are in \AA. 
}
\label{fig4}
\end{figure}

\begin{figure}
\centering{
  \includegraphics[angle=-90,width=0.7\linewidth]{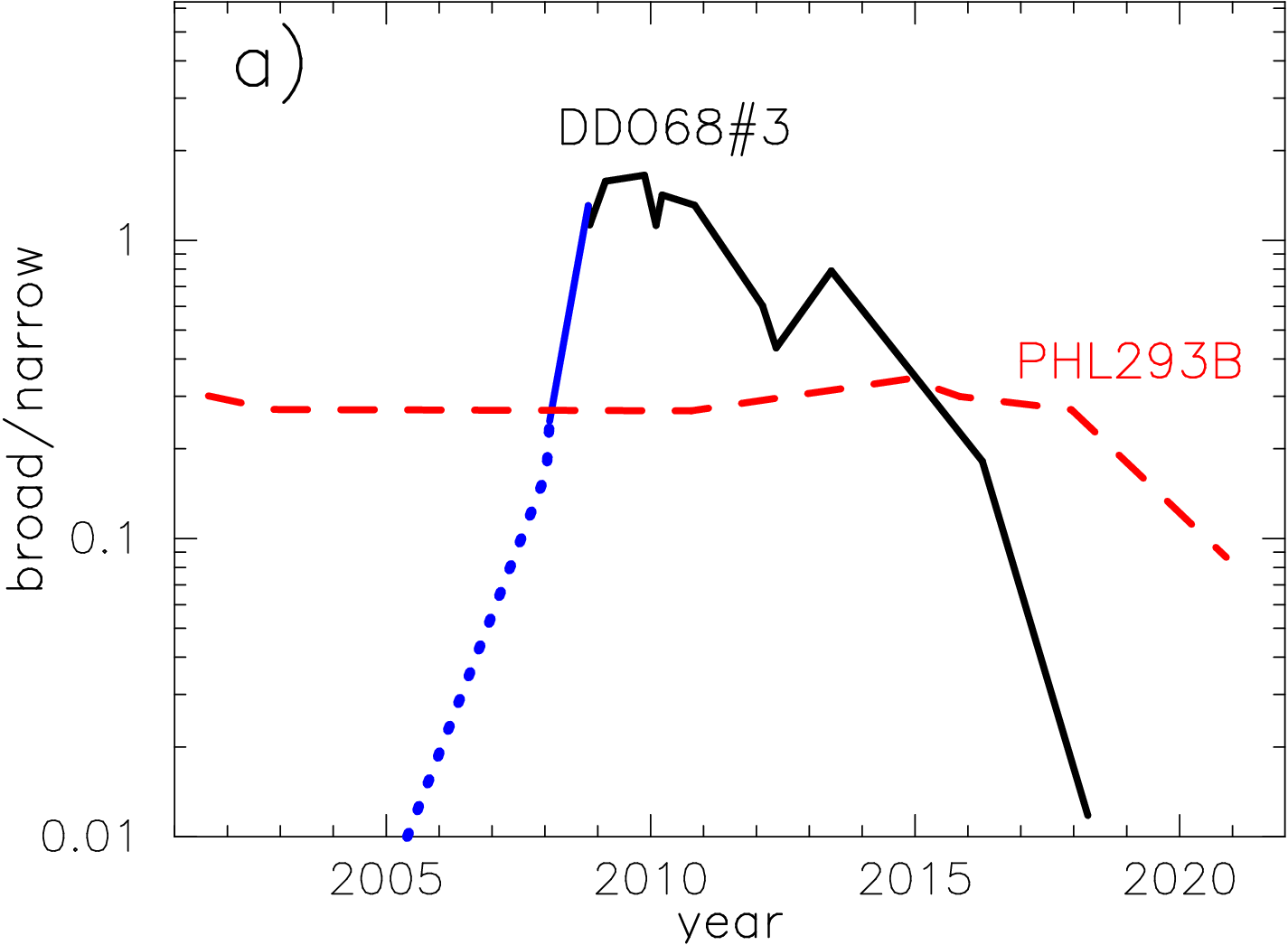}
  \includegraphics[angle=-90,width=0.7\linewidth]{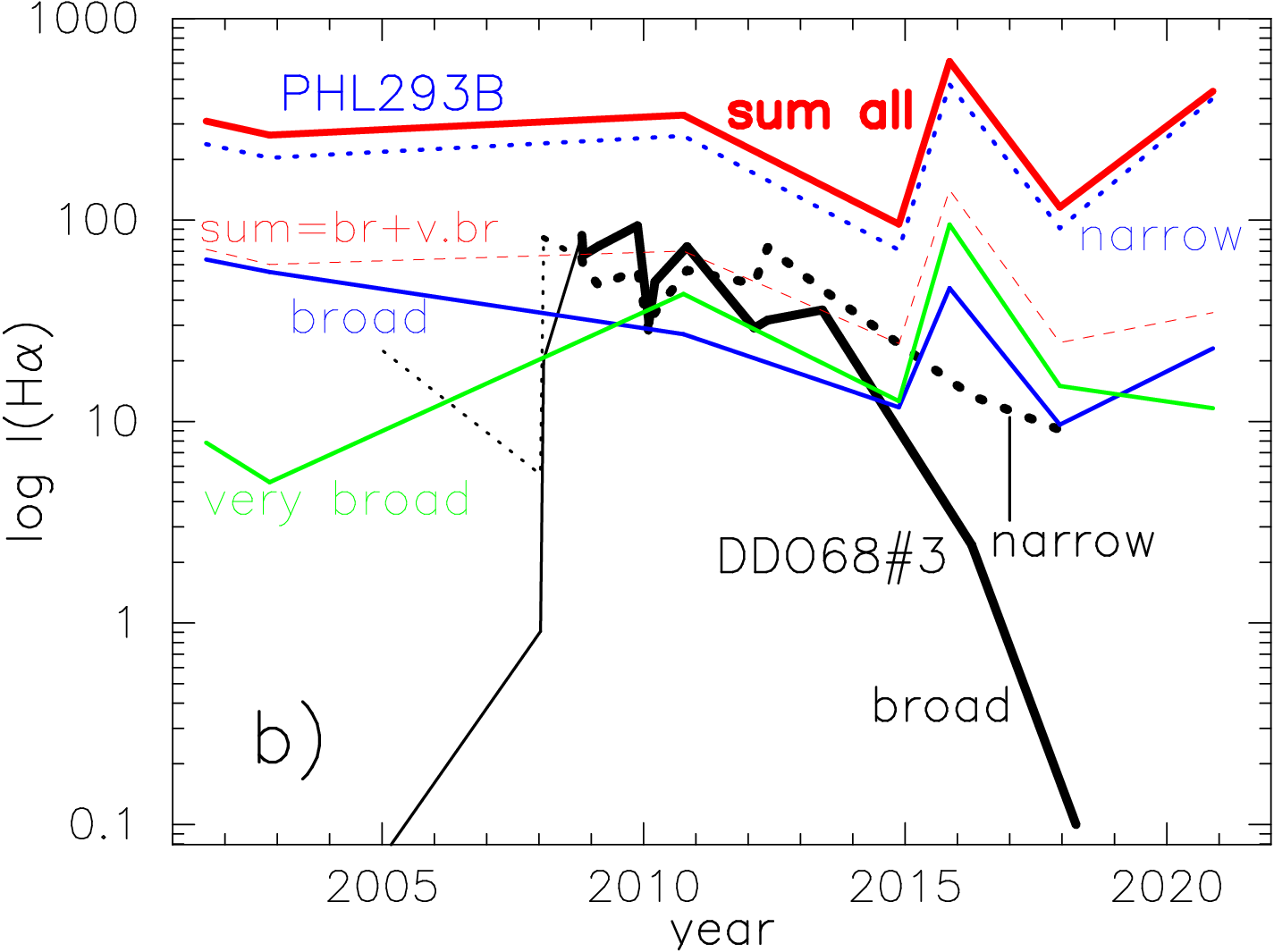}  
}
\caption{a) Temporal variations of the flux ratio of broad-to-narrow H$\alpha$
components in DDO\,68~\#3.  The black solid line shows the APO observations discussed here 
while the blue parts of the curve indicate Special Astrophysical Observatory (SAO) and APO observations by
\citet{Pustilnik2005,Pustilnik2008} (dotted lines) and by
\citet{IzT09} (solid line).
 The temporal variations of the same flux ratio in PHL\,293B is shown by the 
red dashed line. 
  Note that the total flux of the whole broad bump (i.e. the sum of the broad and very
  broad emission components, and of the absorption component when appropriate)
  was taken to be the broad component when calculating this broad-to-narrow ratio. 
  b) Temporal variations of H$\alpha$ fluxes in different  
  components for DDO\,68~\#3. The thick black line shows the APO observations discussed here 
while the thin black line represents data by \citet{Pustilnik2005,Pustilnik2008} and
  \citet{IzT09}. The colored lines show the data for PHL\,293B.
}
\label{fig8}
\end{figure}

\section{OBSERVATIONS AND DATA REDUCTION}\label{sec:OBS}

Over the course of more than a decade, starting in October 2008, we have obtained a series of spectroscopic 
observations for the two SFGs DDO\,68 (its H~{\sc ii} region \#3)
and PHL\,293B, using two different telescopes and instrumental setups.

\subsection{APO Observations}

The first instrumental setup consisted of the Dual Imaging Spectrograph
(DIS) mounted on the ARC 3.5m Apache Point Observatory (APO) telescope.
The blue and red channels of the DIS permitted to simultaneously observe the objects over
a wide range of wavelengths. A long slit giving medium resolution ($R$ = 5000)
was used during all APO observations.
 The  B1200 grating giving medium resolution ($R$ = 5000) with a central wavelength of $\sim$4800 \AA\ and a linear
dispersion of 0.62 \AA\ pixel$^{-1}$ was used in the blue range.
   In the red range, we employ the R1200 grating with a central wavelength of $\sim$6600 \AA\
and a linear dispersion of 0.56 \AA\ pixel$^{-1}$.
   In this way, APO medium resolution spectra of the two SFGs 
DDO\,68~\#3 and PHL\,293B were obtained, 
that span two
wavelength ranges, from $\sim$4150 to 5400 \AA\ in the blue range, and
from $\sim$6000 to 7200\AA\ in the red range. 

The journal of the APO observations, giving the observation dates, the exposure times, the slit widths and the airmasses, 
are shown in Table~\ref{tab6}.

\subsection{LBT Observations}

For PHL\,293B, one high signal-to-noise ratio spectroscopic observation was also carried out
with the 2$\times$8.4m  Large Binocular Telescope (LBT).   
We employ the LBT in the binocular mode utilizing both the MODS1 and MODS2
spectrographs,
equipped with 8022$\times$3088 pixel CCDs.
   Two gratings, one in the blue range, G400L, with a dispersion of
0.5\AA\ pixel$^{-1}$, and another in the red range, G670L, with a dispersion of
0.8\AA\ pixel$^{-1}$, were used.
   The PHL\,293B spectrum was obtained in the wavelength range
$\sim$3150 -- 9500 \AA\ with a 60$\times$1.2 arcsec slit, 
giving a resolving power $R$ $\sim$ 2000.
    The seeing during the observation was in the
range 0.5 -- 0.6 arcsec.
     Three subexposures were derived, resulting in an effective
exposure time of 2$\times$2700 s when adding the fluxes obtained with both
spectrographs, MODS1 and MODS2.
The spectrum of the spectrophotometric standard star GD~71,
obtained during the LBT observation with a 5 arcsec wide slit, was used for flux
calibration. It was also used to correct the red part of the PHL\,293B
spectrum for telluric absorption lines. The journal of the LBT observation is also shown in Table~\ref{tab6}.

\subsection{Data reduction}

  The two-dimensional APO spectra were bias- and flat-field corrected, fixed for
distortion and tilt of frame and background subtracted using
{\sc iraf} routines.
For the LBT observation, the MODS Basic CCD Reduction package
{\sc mods}CCDR{\sc ed} \citep{Pogge2019}
was used for flat field correction and bias subtraction.
Wavelength calibrations of both LBT and APO observations were performed using
spectra of comparison lamps obtained every night before and after the observations.
Each two-dimensional spectrum was aligned along the brightest part of the
galaxy. After wavelength and flux calibration and removal of cosmic particle
trails, all subexposures were summed.
One-dimensional spectra were then extracted along the spatial axis so that the
entire bright part of the H~{\sc ii} region falls into the selected aperture.

 In summary, we have obtained twelve new APO observations of DDO\,68~\#3 and four new APO observations of PHL\,293B, 
together with one new LBT spectrum of PHL\,293B. To increase the time baseline for PHL\,293B, we have also 
included in our analysis the Sloan Digital Sky Survey (SDSS) spectrum obtained with the 2.5m APO telescope on 
22 August 2001 and available in the SDSS archive. This brings the total number 
of PHL\,293B spectra to be analyzed to six. 
More details are given in Tables~\ref{tab7-DDO-G} and \ref{tab7-PHL-G}. The spectra are
shown in Fig.~\ref{fig6} and Fig.~\ref{fig7}, respectively.
All spectra are displayed in the wavelength range
around H$\alpha$ to better emphasize the temporal changes of this line.   
In Fig.~\ref{fig6a} we have also shown two spectra of PHL\,293B over the whole
wavelength range of observations, and that are most separated in time.

 \section{RESULTS}\label{sec:Results}

\subsection{Profile decomposition}

    The most remarkable features in the DDO\,68~\#3 and PHL\,293B spectra
are the strong broad components with P Cygni profiles underlying 
the narrow nebular emission of the hydrogen and helium lines.
For derive quantitative measurements of these features, we have   
decomposed the profiles of the hydrogen emission lines into the sum of  
several Gaussian profiles: a high-intensity 
narrow component, and low-intensity broad and very broad (the latter when needed) 
components, using the {\sc iraf/splot} deblending routine.

The fluxes and full widths at half maximum (FWHM) of the H$\alpha$ narrow and
broad components, along with the terminal velocities and the total luminosities of the broad
components in the two SFGs 
are given in Table~\ref{tab7-DDO-G}
and Table~\ref{tab7-PHL-G} for all spectra shown in Figs.\ref{fig6} and
\ref{fig7}.
The data for the very broad components in PHL\,293B are also given in
Table~\ref{tab7-PHL-G}. Details of the profile decomposition of the H$\alpha$ line can be seen in
Figs.~\ref{fig2},~\ref{fig3},~\ref{fig4}.
    For luminosity determination, the observed fluxes of narrow and broad
emission lines were corrected for the extinction and underlying stellar
absorption taken from \citet{IzT09}, derived in accordance with the prescription of \citet{IzThLip1994}.
Note that the extinction coefficient $C$(H$\beta$) and equivalent width of
the underlying absorption (EW$_{\rm abs}$) derived from the high signal-to-noise ratio MMT spectrum of
DDO\,68~\#3 and the VLT spectrum of PHL\,293B \citep[see table 3 in ][]{IzT09}
are very small in both galaxies. $C$(H$\beta$)
and EW$_{\rm abs}$ have zero values in DDO\,68~\#3 and are respectively equal to 0.08 and 0.05 in PHL\,293B. 

\subsection{DDO\,68~\#3}\label{subsec:DDO68}

   DDO\,68 (UGC 05340) is one of the most metal-deficient SFG known 
\citep[12 + logO/H = 7.15$\pm$0.04, 7.14$\pm$0.07, ][]{IzT09,Annibali2019}.
It is likely in the process of forming by hierarchical merging  \citep{Pustilnik2017,Annibali2019a}. 
Broad spectral features with blueshifted absorption
in the lines of the hydrogen series as well as in some He~{\sc i} lines were first noticed in DDO\,68 by
\citet{Pustilnik2008}. Those authors attributed the broad features
 to the outburst of a LBV located in one of the 
H~{\sc ii} region of  DDO\,68 named Knot 3, and which we will designate by \#3 in
the remainder of this paper.

   Based on photometric and spectroscopic observations, \citet{IzT09} dated the start of 
the strong LBV outburst in DDO\,68~\#3 to be between 2007 February and
2008 January.
   They confirmed the presence of P Cygni profiles in both the H~{\sc i} and
He~{\sc i} emission lines and found that the Fe {\sc ii} emission lines are not
present, in contrast to ``typical'' LBVs \citep{Pustilnik2008,IzT09}.
The absence of Fe {\sc ii} could be due to extremely low metallicity and thus low optical depth precluding considerable radiative pumping.
\citet{Pustilnik2017} described the state of the LBV in 2015-2016 as being in a
fading phase. Observations carried out by \citet{Annibali2019} with LBT/MODS1 and 
LBT/MODS2 in February 2017 do not show the characteristic signs of a LBV star. This
led them to conclude that, by 2017, the LBV was back in a quiescent phase. 
 
  We provide here a more complete picture of the time evolution of the LBV in DDO\,68~\#3
by monitoring its spectrum. 
Our observations start from 2008, and occur as often and as regularly as possible afterwards, ending in 2018.
We show in Fig.~\ref{fig6} the time evolution of the H$\alpha$ line.  
We emphasize that all spectra were obtained with the same telescope and instrumental setup 
(3.5m APO/DIS) and reduced in an uniform manner, so they are directly comparable.
We remark that, as seen in Fig.~\ref{fig6} and more clearly in
Fig.~\ref{fig2}, the [N~{\sc ii}] emission lines near H$\alpha$ are nearly
not detected, due to the low metallicity of DDO\,68~\#3.

Our new APO observations in the monitoring series of DDO\,68~\#3 reveal
that the fluxes of the H$\alpha$ broad components are nearly an
order of magnitude higher at the end of October 2008 as compared to
January 2008, when the LBV was discovered by \citet{Pustilnik2008} (Table~\ref{tab7-DDO-G} and Fig.~\ref{fig8}b, thick solid black line).
The MMT observation of DDO\,68~\#3 on March 2008 of \citet{IzT09} is consistent with that trend. It showed a broad H$\beta$ flux $\sim$2 times lower than the one in the first observation in
our APO monitoring series starting 6 months later, in October 2008.

Our more than a decade APO monitoring showed that the features characteristic of
eruption in the LBV in DDO\,68~\#3 persist until the period somewhere between the end of 2010 and the beginning of 2012, with small variations in the component fluxes. 
   The width of the broad component remains unchanged until October 2010 with
a FWHM $\sim$1000--1200 km s$^{-1}$ (Table~\ref{tab7-DDO-G}).
After 2011 the flux and the velocity of the broad component markedly 
decreased.
Furthermore, the absorption feature in the P Cygni profile disappeared toward the beginning of 2012. 
That disappearance together with the sharp decrease of the 
fluxes of the broad features are possibly due to the dense stellar 
envelope being destroyed and scattered.

   In addition to the presence of broad components with absorption components, observed during the period from 2008 up to 2012 (Fig.~\ref{fig2} a -- g), 
a low-intensity high-velocity redward tail made its appearance in 2012. It 
can be seen as a flux excess above the Gaussian fit to the broad component
(Fig.~\ref{fig2}h and \ref{fig2}i).
  It should be noted, however, that the best fit to the 2012 profiles could
only be obtained by including 
two Gaussians for the broad component (Fig.~\ref{fig3}, a broad one and a
very broad one).
  Thus, a supplementary broad component with a small FWHM = 300 -- 500 km s$^{-1}$ 
is present, in addition to the very broad component with a larger FWHM $\sim$ 1000 km s$^{-1}$
(see Fig.~\ref{fig3}, solid turquoise curves).
  Note the blueward shift in the broad components in Fig.~\ref{fig3} (solid turquoise curves).
  Likely, the appearance of the high-velocity tail and the subsequent disappearance of the P Cygni profile
can be interpreted as the break-up of a dense circumstellar envelope by a rapid outflow.
  In 2013, the spectrum, despite being somewhat noisy, still clearly shows the broad blueward component 
(Fig.~\ref{fig2}j).  
By 2018, the 
broad components have completely gone from the spectra,
suggesting that the LBV shell has disappeared. The outburst event has thus lasted from 2008 to 2013, for a duration of about 5 years. 

   Assuming that the broad components of the emission lines belong exclusively
to the LBV, and that the narrow emission components are predominantly nebular, we
can trace the decay of the LBV outburst by using the flux ratio of the
broad-to-narrow components in each of the hydrogen and helium lines.
   This can be done most reliably for the brightest H$\alpha$ line.
   The temporal evolution of the H$\alpha$ broad-to-narrow flux ratio is
   shown in Fig.~\ref{fig8}a.
   A maximum for this flux ratio is seen between 2009 and 2010. It is more
pronounced than the maximum for the flux of the broad
component (Fig.~\ref{fig8}b).
However, in general, the variation of the broad-to-narrow flux ratio follows 
tightly the flux temporal evolution of the broad component.

To derive line flux luminosities for DDO\,68, we need to know the distance of the dwarf galaxy. 
Based on the same imaging {\sl HST} observations of the dwarf galaxy (GO 11578, PI: Aloisi)
\citet{Cannon2014,Sacchi2016,Makarov2017} have applied the TRGB (Tip of the Red
Giant Branch) method to derive 
distances $D$ = 12.74, 12.65 and 12.75 Mpc for DDO\,68, respectively.    
New {\sl HST} observations of a stream-like system associated with DDO\,68
by \citet{Annibali2019a} gives $D$ = 12.8 $\pm$ 0.7 Mpc, using the same TRGB
method. The mean of these values is $D$ = 12.7 Mpc, which we adopt. 
This distance is more than twice as large than the 
 previous indirect distance determination of 
 $D$$\sim$5 -- 7 Mpc (NED) \citep[see e.g. ][]{Pustilnik2005,IzT09}.

With this distance, the luminosity of the H$\alpha$ broad component at the 
maximum is $\sim$ 2$\times$10$^{38}$ ergs s$^{-1}$. 
This maximum luminosity and the FWHM ($\sim$1000--1200 km s$^{-1}$) of the broad component of the lines during 
the transient event are in the ranges of those observed for
LBV stars. The data thus suggest that a LBV star in the H~{\sc ii} region DDO\,68~\#3 has undergone an outburst during the period 2008--2013.

   The terminal velocity $v$$_{\rm term}$ of the LBV stellar wind
is an important parameter for confronting theory with observation.
It is obtained from the
wavelength difference between the maximum of the broad emission line profile
and the minimum of the blue absorption line profile. The terminal velocity 
    $v$$_{\rm term}$ in DDO\,68~\#3 is of $\sim$800 km s$^{-1}$ and
does not change significantly over time (Table~\ref{tab7-DDO-G}).
    It remains also  unchanged during the ``maximum'' of the eruption, as
well as during the decay period starting in 2013. 
    Nearly the same value of the expanding wind terminal velocity was reported by \citet{IzT09} and  \citet{Pustilnik2017}.

To summarize, the new data suggest that the broad component fluxes of the hydrogen lines
of the LBV reached a maximum during the time interval $\sim$ 2008--2012. 
The narrow component also reached a maximum during the same period, but less pronounced.
 The broad-to-narrow component flux ratio 
reached a maximum at about the same period.
This also supports the hypothesis that the LBV star
underwent an eruptive event. It can be seen in Fig.~\ref{fig8} that the H$\alpha$ luminosity of 
the LBV star is now at a minimum or close to it.

\subsection{PHL\,293B}\label{subsec:PHL 293B}

The situation with understanding the physical processes taking place in PHL\,293B is
more complex than in DDO\,68~\#3.
PHL\,293B (J2230--0006) is a well known SFG with a moderately low metallicity
12+logO/H $\sim$ 7.6--7.7 
\citep[e.g. ][]{Papaderos2008,IzGusFrHenkel2011,Fernandez2018}.
  \citet{IzT09} first detected a broad 
component in the strong hydrogen emission lines with P Cygni profiles of PHL\,293B
in a VLT
UVES spectrum obtained on 8 November 2002, and in a SDSS spectrum taken on 22 August
2001. 
Those authors suggested that these broad features are due to a transient LBV phenomenon. Earlier observations of PHL\,293B by \citet{Kinman1965} and
\citet{French1980} did not mention any broad emission.
The broad component was again detected by \citet{IzGusFrHenkel2011} in a 
VLT X-Shooter observation on 16 Aug 2009.
Later observations have been carried out, and
various other hypotheses have been proposed to interpret them.
Thus, \citet{Terlevich2014} have suggested that the observations of PHL\,293B can be explained by 
a young dense expanding supershell driven by a stellar cluster wind and/or two supernova remnants
or by a stationary wind. Hydrodynamic models 
have been built by \citet{Tenorio2015}.

On the basis of 10.4m GTC (Gran Telescopio Canarias)/MEGARA observations performed in July 2017,
\citet{Kehrig2020} found a flux ratio $I$$_{br}$/$I$$_{nar}$ $\sim$0.10 in the H$\alpha$ emission
line of all integrated regions. They 
 interpret this low value as due to a diminution 
 of the broad H$\alpha$ emission.

\citet{Allan2020}, from spectroscopic observations including the 2019 VLT X-Shooter
data and radiation transfer modeling, report the absence of
the broad emission component since 2011 and conclude that the LBV was in an
eruptive state during 2001 - 2011 that has ended.
\citet{Burke2021} also report the decrease of the $I$$_{br}$/$I$$_{nar}$ ratio from 0.41
(SDSS, 2001) to $\sim$0.10, using Gemini observations taken in December 2019. 
Despite the fact that a 
AGN-like damped random walk model works well to fit the observed light curve,  those authors 
concluded that a long-lived stellar transient of type SN IIn  can better
explain all the data for PHL\,293B.

On the other hand, 
\citet{Prestwich2013} have emphasized the lack of
X-ray emission in the galaxy, establishing an 
upper limit of $L_X$ $\sim$ 3 $\times$ 10$^{38}$ erg s$^{-1}$ (their table 6).
This casts some doubt on the supernova model, thought to be the most probable 
one for explaining the PHL 293
phenomenon. \citet{Kehrig2020} have detected P Cygni profiles in
the H$\alpha$ and H$\beta$ emission lines in July 2017, while those features were not found by \citet{Burke2020} in
H$\alpha$ in December 2019. These variations suggest that 
long-time monitoring, such as the observations described here, is important for distinguishing between various models.

Contrary to the conclusions of \citet{IzT09}, based on the archival VLT/UVES
observations on 2002-11-08, that P Cygni profiles are seen only in the hydrogen
lines, and those of \citet{Terlevich2014}, our new PHL\,293B observations,
as well as a reconsideration of the SDSS archival data (2001-08-22) and the VLT/X-Shooter
observation by \citet{IzGusFrHenkel2011} on 2009-08-16, show that
blue-shifted absorption lines are detected not only in hydrogen emission lines,
but also in He~{\sc i} lines (Fig.~\ref{fig6a}). 
In other words, the situation is the same as in the
case of DDO\,68~\#3 which has a much lower metallicity.
However, the permitted Fe {\sc ii} emission lines, which generally originate in dense circumstellar
envelopes and are usually seen in the spectra of LBV stars experiencing a giant
eruption and creating an envelope, have not been detected in our new
observations.
 Note that \citet{Terlevich2014} found blue-shifted Fe {\sc ii} absorption with
a terminal velocity of $\sim$ 800 km s$^{-1}$.
We did not detect permitted Fe~{\sc ii} lines in the LBT spectrum obtained in
2020. Only forbidden [Fe~{\sc iii}]$\lambda$$\lambda$4658, 4702, 4755:, 4986,
5270\AA\ emission lines were found in this spectrum, with no absorption features. 

In the decomposition of the strong emission lines, we have always attempted to fit the broad
component in the simplest way possible, i.e., with the smallest number
(preferably one) of Gaussian profiles. 
 We also tried to perform fitting with Lorentzian profiles. But the results were always worse than in the case of Gaussian profiles.
In the case of
PHL293B, not one but two broad components are required to match the observed
profiles well. The decomposition of H$\alpha$ for all spectra from
Table~\ref{tab7-PHL-G} into narrow, broad, very broad (and absorption lines,
whenever warranted), are presented in Fig.~\ref{fig4}.

The P Cygni profile has persisted over some two decades, during the period 2001--2020, 
as is clearly seen in Fig.~\ref{fig6a}. 
    The blueward absorption feature with wavelength $\sim$6545\AA\ is close to
    the nitrogen emission line [N {\sc ii}]$\lambda$6548\AA\
so that in many
cases decomposition of the absorption profile in H$\alpha$ emission line is difficult. The absorption is likely present, but
masked by strong broad and very broad components and the nitrogen emission line
(Fig.~\ref{fig7} and Fig.~\ref{fig4}).

  We obtain a high terminal velocity $v_{term}$$\sim$800 km s$^{-1}$ for the absorption
component in the Balmer lines. This value is 
the same as the ones derived by \citet{IzT09}, \citet{IzGusFrHenkel2011} and
\citet{Terlevich2014} from X-Shooter observations made in 
Aug. 2009 and in Aug.-Sep. 2009, respectively.
The luminosity of the broad bump (i.e. the sum of broad and very broad
components) varies from a few 10$^{38}$ erg s$^{-1}$ up to
$\sim$10$^{39}$ erg s$^{-1}$.

The very broad component derived by us from the SDSS spectrum taken in August 2001
(Fig.~\ref{fig4}, Table~\ref{tab7-PHL-G}) has nearly the same
redshift as \citet{Terlevich2014}'s faint ultra-broad component measured from 
their 2009 observations and shown in their figure 2  but a flux that is $\sim$100 times
lower. Note that \citet{Terlevich2014} also fitted H$\alpha$ in their X-Shooter
observation by two broad components.
At about the same redshift, \citet{IzGusFrHenkel2011} also saw an excess in their fit of the 
H$\beta$ broad component (their figure 3). 
In our subsequent series of observations of PHL\,293B starting in 2010, the very broad
component does not show any redward shift. On the contrary, the broad and very broad
components are centered practically at the systemic velocity of the galaxy.
This means that, at least since 2010, there has been no large velocity outflows.
We remark also that the broad components in the SDSS spectrum dated 2001 \citep{IzT09}, and
 in the VLT X-Shooter spectra dated 2009
\citep{IzGusFrHenkel2011,Terlevich2014}, have quite large FWHMs,
$\sim$ 1500,
1000 and 1000 km s$^{-1}$, respectively. After 2010, these components become 
much narrower, with FWHM $\sim$ 160 -- 180 km s$^{-1}$.
  This signifies the fact that the 
velocity dispersions
 of the moving and radiating matter have decreased, at least in the broad components.

 If the ionization parameter is close to the limiting value of log$U$ $\la$--2,
the radiation pressure can prevail over the ionized gas pressure
\citep{Dopita2002}, and some contribution of a radiation-driven (i.e. by LyC and/or
Ly$\alpha$) superwind from a young stellar cluster, producing a high-velocity very broad
component, will be possible \citep{Komarova2021}.
To check this possibility, we use accurate oxygen abundance and an
observational indicator of the ionization parameter, O32 = [O~{\sc iii}] 4959,5007/[O~{\sc ii}]3727,
for PHL\,293B
(12 + logO/H = 7.72 and O32 = 15.52 from \citet{IzT09} and 12 + logO/H = 7.71
and O32 = 14.21 from \citet{IzGusFrHenkel2011}) to estimate log$U$. We get
\citep[following ][]{Kobulnicky2004} log$U$ = --2.2 and log$U$ = --2.3,
respectively, which are close to the maximum value, but still below it. 
 Note that it is higher than log$U$ = --2.9 derived from the MMT observation by
\citet{IzT09} for DDO\,68~\#3.

In general, the temporal variations of both the broad-to-narrow component flux
ratios and of the H$\alpha$ fluxes of the narrow and broad components for
PHL\,293B (Fig.~\ref{fig8}, colored lines) are very different from those 
of DDO\,68~\#3 (Fig.~\ref{fig8}, black lines). 
A strong variability of the  H$\alpha$ fluxes, simultaneously in both the narrow
and broad and very broad components, from 2011 to 2018, is seen in PHL\,293B
(Fig.~\ref{fig8}b) .
Flux jumps by a factor of $\sim$6 are observed in both the narrow and the sum of
the broad and very broad components (Table~\ref{tab7-PHL-G}). 
We have checked the logs of the observations and have determined that all observations
were carried out in photometric conditions, so that the flux variations cannot be attributed to 
sky variability. 
 The flux variations, both in the narrow and broad and
even very broad components (a sharp flux decrease in 2014 
followed by a rise at the end of 2015, followed by another decrease in 2017)   
can be explained by inaccurate pointing when observing with a
narrow slit of 0.9 arcsec. 
However the variations are likely to be real because 
the light curves derived by \citet{Burke2021}
from SDSS and DES (Dark Energy Survey) imaging between 1998 and 2018 (their figure 2) also show
small-amplitude variability in the $g$ and $r$ bands during the same years. 
The synchronicity in the variability of fluxes in the narrow and broad
components does not hold anymore starting 2018 
\citep[see also the fading of the broad component in ][]{Allan2020,Burke2021}.

In contrast to the outburst nature of the variability over time of the
broad-to-narrow flux ratio in DDO\,68~\#3, which manifests itself in 
the form of a 
peak (Fig.~\ref{fig8}a, black line), PHL\,293B does not undergo such a type of
variation.
Its broad-to-narrow flux ratios 
depend weakly on time, at least over the past two decades. 
However, a decrease in the broad-to-narrow flux ratio can be seen
starting from the very end of 2017 and continuing to 2020
(Fig.~\ref{fig8}a). This decrease can be partly explained by an increase of the narrow component relative to the broad one (Fig.~\ref{fig8}b, blue and red dashed and green solid
lines).

In the case of PHL\,293B, the behavior of the broad component may be
explained by several effects. 
Enduring broad hydrogen Balmer P Cygni profiles with absorption feature
blue-shifted by $\sim$800 km s$^{-1}$, can indicate the presence of fast moving ejecta.
The persistent luminosity of the broad Balmer emission  
and the possibility that the Balmer emission could be due to a long-lived
stellar transient, like LBV/SN, motivate additional follow-up spectroscopy to distinguish between these effects.

\subsection{LBVs in other SFGs}

We now compare the behavior of the candidate LBVs in DDO\,68~\#3 and PHL\,293B with those in other SFGs.

The LBV in the relatively low-metallicity  galaxy Mrk 177
= UGCA 239 at the post-merger stage, with 12 + logO/H = 8.58 and $D_L$ = 28.9 Mpc, 
has undergone multiple outbursts during the last 20 years \citep{Kokubo2021}.
The luminosity of the broad H$\alpha$ component of Mrk 177 varies from a maximum of 
10$^{40}$ erg s$^{-1}$ during the strongest explosion to
10$^{39}$ erg s$^{-1}$ during the next two strong explosions.
This is an order of magnitude higher than the corresponding values  
of $L$(H$\alpha$) (broad) = (2 -- 9)$\times$10$^{38}$ erg s$^{-1}$ in PHL\,293B. As for the H$\alpha$
broad-to-narrow flux ratio for the LBV in Mrk 177, it varies in the interval 3.4 -- 4.5
during last two outbursts. This is in the same range as the LBV outburst maximum in
DDO\,68~\#3, but one order of magnitude higher than in PHL\,293B.
The strongest outburst in Mrk 177 during 2001 is characterised by a H$\alpha$
broad-to-narrow ratio that is approximately 10 times higher than those during
subsequent explosions, i.e. $\sim$ two orders of magnitude higher than in
PHL\,293B.  

On the other hand,  the luminosity $L$(H$\alpha$) $\sim$10$^{38}$ erg s$^{-1}$ \citep{Drissen2001}
of the LBV-V1 in the low-metallicity \citep[12 + logO/H = 7.89, ][]{Iz97}
galaxy Mrk 71 (NGC 2363A)  is 
lower than that of the cLBV in PHL\,293B, with a similar metallicity.

 \section{CONCLUSIONS}\label{sec:conclusion}

Over nearly two decades,  we have monitored the time variation of the
broad component fluxes and of the broad-to-narrow flux ratios of the strong hydrogen
and helium emission lines in two low metallicity star-forming galaxies (SFG), DDO\,68 (12 + logO/H = 7.15) and
PHL\,293B (12 + logO/H = 7.72). These two SFGs have the particularity of harboring candidate luminous blue variable (cLBV) stars.
We have carried out this monitoring by obtaining over time spectra of these two objects with the 
3.5m Apache Point Observatory (APO)
telescope and the 2$\times$8.4m Large Binocular Telescope (LBT). 
Our main results are the following:
     
(1) The broad emission with a P Cygni profile of the H$\alpha$ line emitted by the DDO\,68 H~{\sc ii} 
region ~\#3 shows a marked increase of its
luminosity during the period 2005 to 2017, 
reaching a maximum 
$L$(H$\alpha$) of $\sim$ 
10$^{38}$ erg s$^{-1}$ in 2008 -- 2011, adopting the distance derived from brightness of the Tip of the Red Giant branch (TRGB).
The absorption feature of the P Cygni-like profile and the
broad component rapidly decayed and disappeared in 2018. 
These properties are characteristic of an eruptive event in a LBV star.
The derived H$\alpha$ luminosity and the FWHM $\sim$1000--1200 km s$^{-1}$ of the broad
component during the eruption event are in the range of those observed for LBV
stars.
A terminal velocity $v$$_{\rm term}$$\sim$800 km s$^{-1}$ is measured from the absorption profile. 
It does not change significantly over
time. These observations 
are also consistent with the earlier findings 
of \citet{Pustilnik2017} who described the LBV in DDO\,68~\#3 as going through a fading phase during
2015--2016, and with those of \citet{Annibali2019} who did not find any 
characteristic sign of a LBV star at the beginning of 2017. 
Thus, our spectroscopic monitoring indicates that the LBV has passed through the
maximum of its eruption activity and is now in a fairly quiet phase. 
  
(2) The situation is quite different in the BCD PHL\,293B. Since the discovery of the cLBV in it on the basis of SDSS 2001 observations, the fluxes of the broad 
component and the broad-to-narrow flux ratios of the hydrogen 
emission lines in PHL\,293B have remained nearly constant,  with small variations, over 16 years, at least until 2015.
A decrease  of the broad-to-narrow flux ratio in recent years (2017--2020)
can partly be attributed to an increase of the narrow component flux relative to that of the broad
component.
The luminosity of the broad H$\alpha$ component varies from
$\sim$2$\times$10$^{38}$ erg s$^{-1}$ to $\sim$10$^{39}$ erg s$^{-1}$, and the FWHM  varies in the range 
$\sim$500--1500 km s$^{-1}$ over the whole period of monitoring of PHL\,293B, from 2001 to 2020.

Unusually persistent P Cygni features with broad and very broad components of
hydrogen 
emission lines and blueward absorption in the H~{\sc i} and He~{\sc i} emission
lines are clearly visible, even at the very end of 2020, despite the several
decreases of the broad-to-narrow flux ratio in the most recent years (2017-2020).
A terminal velocity $v$$_{\rm term}$$\sim$800 km s$^{-1}$ is measured in PHL\,293B, similar to the one obtained for DDO\,68~\#3, although the latter is 3.7 times more metal deficient than the former. 
The terminal velocity does not change significantly with 
time. 
The near-constancy of the H$\alpha$ flux suggests that the cLBV in 
PHL\,293B is a long-lived stellar transient of type LBV/SN IIn.
However,  other mechanisms such as a stationary wind from a young stellar cluster, 
cannot be ruled out. Further spectroscopic time monitoring of  the BCD is needed to narrow down 
the nature of the phenomenon in PHL\,293B.

\section*{Acknowledgements}

N.G.G. and Y.I.I. acknowledge support from
the National Academy of Sciences of Ukraine (Project No. 0121U109612
``Dynamics of particles and collective excitations in high energy physics,
astrophysics and quantum microsystem''). T.X.T. thanks the hospitality of the 
Institut d'Astrophysique in Paris where part of this work was carried out.
  The APO 3.5 m telescope is owned and operated by the Astrophysical
Research Consortium (ARC).
   The LBT is an international collaboration among institutions in the United
States, 
Italy and Germany. LBT Corporation partners are: The University of Arizona on 
behalf of the Arizona university system; Istituto Nazionale di Astrofisica, 
Italy; LBT Beteiligungsgesellschaft, Germany, representing the Max-Planck 
Society, 
the Astrophysical Institute Potsdam, and Heidelberg University; The Ohio State 
University, and The Research Corporation, on behalf of The University of Notre 
Dame, University of Minnesota and University of Virginia.
   This paper used data obtained with the MODS
spectrographs built with funding from NSF grant AST-9987045 and the NSF
Telescope System Instrumentation Program (TSIP), with additional funds from
the Ohio Board of Regents and the Ohio State University Office of Research.
   {\sc iraf} is distributed by the 
National Optical Astronomy Observatories, which are operated by the Association
of Universities for Research in Astronomy, Inc., under cooperative agreement 
with the National Science Foundation.
  Funding for the Sloan Digital Sky Survey IV has been provided by
the Alfred P. Sloan Foundation, the U.S. Department of Energy Office of
Science, and the Participating Institutions. SDSS-IV acknowledges
support and resources from the Center for High-Performance Computing at
the University of Utah. The SDSS web site is www.sdss.org.
SDSS-IV is managed by the Astrophysical Research Consortium for the 
Participating Institutions of the SDSS Collaboration. 
This research has made use of the NASA/IPAC Extragalactic Database (NED), which 
is operated by the Jet Propulsion Laboratory, California Institute of 
Technology, under contract with the National Aeronautics and Space 
Administration.

\section{DATA AVAILABILITY}

The data underlying this article will be shared on reasonable request to the corresponding author.

\bsp

\label{lastpage}


\begin{thebibliography}{}


\bibitem[Allan et al.(2020)]{Allan2020} Allan A. P., Groh J. H., Mehner A.,
  Smith N., Boian I., Farrell E. J., Andrews J., 2020, \mnras, 496, 1902  

\bibitem[Annibali et al.(2019a)]{Annibali2019} Annibali F., La Torre V., Tosi M.  et al., 2019a, \mnras, 482, 3892 

\bibitem[Annibali et al.(2019b)]{Annibali2019a} Annibali F., Bellazzini M., Correnti M.  et al., 2019b, \apj, 883, 19 

\bibitem[Burke et al.(2020)]{Burke2020} Burke C. J., Baldassare V. F., Liu X.  et al., 2020,  \apjl, 894, L5 

\bibitem[Burke et al.(2021)]{Burke2021} Burke C. J., Liu X., Chen Y.-C.,
  Shen Y., \& Guo H., 2021, \mnras, 504, 543 

\bibitem[Cannon et al.(2014)]{Cannon2014} Cannon J. M., Johnson M., McQuinn K. B. W. et al., 2014, \apj, 787, L1

\bibitem[Crowther(2007)]{Crowther2007} Crowther P. A., 2007, \araa, 45, 177

\bibitem[Davidson \& Humphreys(1997)]{DavidsonHumphreys1997}Davidson K.,
  Humphreys R. M., 1997, \araa, 35, 1

\bibitem[Davidson(1999)]{Davidson1999} Davidson K., 1999,
in ``Eta Carinae At The Millennium'', eds. J. A. Morse, R. M. Humphreys, and
A. Damineli, \aspc, 179, 6 

\bibitem[Dopita et al.(2002)]{Dopita2002} Dopita M. A., Groves B. A., Sutherland R. S., Binette L., Cecil G., 2002, \apj, 572, 753 

\bibitem[Drissen et al.(1997)Drissen, Roy \& Robert]{Drissen1997}
  Drissen L., Roy J.-R., Robert C., 1997, \apj, 474, L35

\bibitem[Drissen et al.(2001)]{Drissen2001} Drissen L., Crowther P. A.,
  Smith L. J., Robert C., Roy J.-R., Hillier D. J., 2001, \apj, 546, 484

\bibitem[Fern\'andez et al.(2018)]{Fernandez2018} Fern\'andez V., Terlevich E.,
D\'iaz A., Terlevich R., Rosales-Ortega F. F., 2018, \mnras, 478, 5301

\bibitem[French(1980)]{French1980} French H. B., 1980, \apj, 240, 41

\bibitem[Grassitelli et al.(2020)]{Grassitelli2020} Grassitelli L, Langer N.,
  Mackey J., Gr\"afener G., Grin N. J., Sander A. A. C., Vink J. S.,
  2020, \aap, 647, 99 

\bibitem[Guseva et al.(2000) Guseva, Izotov \& Thuan]{GIT00}
Guseva N. G., Izotov Y. I., Thuan T. X., 2000, \apj, 531, 776 

\bibitem[Guseva et al.(2012)]{Guseva2012} Guseva N. G., Izotov Y. I.,
  Fricke K. J., Henkel C.,
  2012, \aap, 541, A115  
  
\bibitem[Humphreys \& Davidson(1994)]{HumphreysDavidson1994} Humphreys R. M.,
Davidson K., 1994, \pasp, 106, 1025 

\bibitem[Humphreys et al.(2013)]{Humphreys2013} Humphreys R. M., Davidson K.,
Grammer S., Kneeland N., Martin J. C., Weis K., Burggraf B.,
2013, \apj, 773, 46

\bibitem[Humphreys et al.(2017)]{Humphreys2017} Humphreys R. M., Davidson K.,
Hahn D., Martin J. C., Weis K., 2017, \apj, 844, 40

\bibitem[Humphreys (2019)]{Humphreys2019} Humphreys R. M., 2019, \gal, 7, 75

\bibitem[Izotov et al.(1994)]{IzThLip1994} Izotov Y. I., Thuan T. X.,
Lipovetsky V A., 1994, \apj, 435, 647

\bibitem[Izotov et al.(1997)Izotov, Thuan \& Lipovetsky]{Iz97} Izotov Y. I.,
Thuan T. X., Lipovetsky V. A., 1997, \apjs, 108, 1 

\bibitem[Izotov et al.(2007)Izotov, Thuan \& Guseva]{IzTG07}
Izotov Y. I., Thuan T. X., Guseva N. G., 2007, \apj, 671, 1297 

\bibitem[Izotov \& Thuan(2008)]{IzT08}
Izotov Y. I., Thuan T. X., 2008, \apj, 687, 133 

\bibitem[Izotov \& Thuan(2009)]{IzT09}
Izotov Y. I., Thuan T. X., 2009, \apj, 690, 1797 

\bibitem[Izotov et al.(2011)]{IzGusFrHenkel2011} Izotov Y. I., Guseva N. G.,
Fricke K. J., Henkel C., 2011, \aap, 533, A25 

\bibitem[Kehrig et al.(2020)]{Kehrig2020} Kehrig C., Iglesias-P\'aramo J.,
  V\'ilchez J. M. et al. 2020, \mnras, 498, 1638 

\bibitem[Kinman(1965)]{Kinman1965} Kinman T. D., 1965, \apj, 142, 1241

\bibitem[Kobulnicky \& Kewley(2004)]{Kobulnicky2004} Kobulnicky H. A., Kewley L. J., 2004, \apj, 617, 240

\bibitem[Kokubo (2021)]{Kokubo2021} Kokubo M., 2021, \mnras, in press;
  arXiv:2101.07797 

\bibitem[Komarova et al.(2021)]{Komarova2021} Komarova L., Oey M. S.,
Krumholz M. R., Silich S., Kumari N., James B. L., 2021, \apj, 920, L46
  
\bibitem[Lamers et al.(1983)]{Lamers1983} Lamers H. J. G. L. M., de Groot M.,
Cassatella A., 1983, \aap, 128, 299 

\bibitem[Leitherer et al.(2001)]{Lei01} Leitherer C., Le\=ao J. R. S., 
Heckman T. M., Lennon D. J., Pettini M., Robert C., 2001, \apj, 550, 724

\bibitem[Makarov et al.(2017)]{Makarov2017} Makarov D. I., Makarova L. N.,
  Pustilnik S. A., Borisov S. B., 2017, \mnras, 466, 556

\bibitem[Massey et al.(2000)]{Massey2000} Massey P., Waterhouse E.,
  DeGioia-Eastwood K., 2000, \aj, 119, 2214

\bibitem[Papaderos et al.(2008)]{Papaderos2008} Papaderos P., Guseva N. G.,
Izotov Y. I., Fricke K. J., 2008, \aap, 491, 113

\bibitem[Pogge(2019)]{Pogge2019} Pogge R., 2019,
https://doi.org/10.5281/zenodo.2550741

\bibitem[Prestwich et al.(2013)]{Prestwich2013} Prestwich A. H., Tsantaki M., Zezas A., Jackson F., Roberts T. P., Foltz R., Linden T., Kalogera V.,
2013, \apj, 769, 92  
  
\bibitem[Pustilnik et al.(2005)]{Pustilnik2005} Pustilnik S. A.,
 Kniazev A. Y., Pramskij A. G., 2005, \aap, 443, 91 

\bibitem[Pustilnik et al.(2008)]{Pustilnik2008} Pustilnik S. A.,
Tepliakova A. L., Kniazev A. Y., Burenkov A. N., 2008, \mnras, 388, L24 

\bibitem[Pustilnik et al.(2017)]{Pustilnik2017} Pustilnik S. A., Makarova,
L. N., Perepelitsyna, Y. A., Moiseev A. V., Makarov D. I.,
2017, \mnras, 465, 4985 

\bibitem[Sacchi et al.(2016)]{Sacchi2016} Sacchi E., Annibali F., Cignoni M. et al., 2016, \apj, 830, 3

\bibitem[Schaerer et al.(1999) Schaerer, Contini \& Pindao]{SCP99}
Schaerer D., Contini T., Pindao M., 1999, \aaps, 136, 35

\bibitem[Smith et al.(1994)]{Smith1994} Smith L. J., Crowther P. A.,
  Prinja R. K., 1994, \aap, 281, 833

\bibitem[Smith et al.(2011)]{Smith2011} Smith N., Li W., Silverman J. M.,
  Ganeshalingam M., Filippenko A. V., 2011, \mnras, 415, 773

\bibitem[Sobral et al.(2020)]{Sobral2020} Sobral D., Matthee J., Darvish B. et al., 2018, \mnras, 477, 2817

\bibitem[Solovyeva et al.(2020)]{Solovyeva2020} Solovyeva Y., Vinokurov A., Sarkisyan A. et al., 2020, \mnras, 497, 4834
 
\bibitem[Tenorio-Tagle et al.(2015)]{Tenorio2015} Tenorio-Tagle G., Silich S.,
Mart\'inez-Gonz\'alez S., Terlevich R., Terlevich E., 2015, \apj, 800, 131

\bibitem[Terlevich et al.(2014)]{Terlevich2014} Terlevich R., Terlevich E.,
Bosch G.,  D\'iaz \'A., H\"agele G., Cardaci M., Firpo V.,
2014, \mnras, 445, 1449

\bibitem[Vink(2012)]{Vink2012} Vink J. S., 2012, in ``Eta Carinae and the
supernova imposters'', eds. R. Humphreys and K. Davidson, \assl, 384, 221

\bibitem[Weis \& Bomans(2020)]{Weis2020} Weis K., Bomans D. J., 2020,
\gal, 8, 20

\bibitem[Wofford et al.(2020)]{Wofford2020} Wofford A. et al.,
2020, \mnras, 493, 2410   
  

\end{thebibliography}
\end{document}